\documentclass{aa}
\usepackage[varg]{txfonts}
\usepackage[T1]{fontenc}
\usepackage[utf8]{inputenc}
\usepackage{textcomp} 
\usepackage{gensymb}
\usepackage{natbib}
\usepackage{amsmath}
\usepackage{graphicx,subfig}

\usepackage[normalem]{ulem}

\fancypagestyle{firstpage}{%
	\fancyhf{}%
	\fancyfoot[RO]{\thepage}%
	\fancyfoot[LE]{\thepage}%
}

\begin{document}

\titlerunning{Evolution of a migrating planet with a binary companion}

\title{Evolution of a migrating giant planet in the presence of an inclined binary companion}
\author{A. Roisin\inst{1} 
        \and A.-S. Libert\inst{1}
}
\institute{naXys, Department of Mathematics, University of Namur, 8 Rempart de la Vierge, B-5000 Namur, Belgium}
\date{Received ... / Accepted ...}
\abstract {} {There are a growing number of giant planets discovered moving around one stellar component of a binary star, most of which have very diverse eccentricity. These discoveries raise the question of their formation and long-term evolution because the stellar companion can strongly affect the planet formation process. We aim to study the dynamical influence of a wide binary companion on the evolution of a single giant planet migrating in a protoplanetary disk.} {Using a symplectic n-body integrator adapted for binary star systems and modeling the dissipation due to the disk by appropriate formulae emerging from hydrodynamical simulations, we carried out 3200 simulations with different orbital parameters for the planet and different eccentricity and inclination values for the binary companion. The long-term evolution of the planets was followed for 100 Myr and the different dynamical behaviors were unveiled using a quadrupolar Hamiltonian approach.}{We show that a capture in a Lidov-Kozai resonant state is far from automatic when the binary companion star is highly inclined, since only $36 \%$ of the systems end up locked in the resonance at the end of the simulations. Nevertheless, in the presence of a highly inclined binary companion, all the planetary evolutions are strongly influenced by the Lidov-Kozai resonance and the nonresonant evolutions present high eccentricity and inclination variations associated with circulation around the Lidov-Kozai islands.}
{}


\maketitle

\section{Introduction}

About half of the Sun-like stars are part of multiple-star systems \citep{Duquennoy_1991,Raghavan_2010}. Since the first discoveries of giant planets in binary stars during the past decade (e.g., GL86 \citep{Queloz_2010}, $\gamma$ Cephei \citep{Campbell_1988,Hatzes_2003}), their number continues to increase, and today more than 100 S-type planets (also known as circumprimary planets) are known that move around one stellar component \citep{Schwarz_2016}. Most of these are found in wide binaries with separations larger than 500 AU. Giant planets in wide binaries have slightly higher eccentricities than around single stars \citep{Kaib_2013} and some reported eccentricities are even higher than 0.85. 

The question of the formation and long-term evolution of these planets remains open, since the stellar companion can strongly affect the planet formation process. Extensive studies have analyzed the long-term stability and  habitability of giant planets detected in binary star systems \citep[e.g.,][]{Dvorak_1988,Holman_1999,Haghighipour_2010,Bazso_2017}. Planet formation has principally been investigated for close binaries in which giant planets have been detected despite a very strong perturbation of the stellar companion. For S-type planets in wide binaries, owing to a higher stellar separation, it is reasonable to expect that the binary companion would have a more limited (but still significant) impact on the planet formation process. 

The different stages of planet formation can be affected by a binary companion \cite[see, e.g.,][for a review]{Thebault_2016}. Firstly, during the formation of the protoplanetary disk, the disk can be truncated by the stellar companion \citep{Artymowicz_1994,Savonije_1994}. Observations have shown that disks in close binaries tend to be less frequent and less massive than around single stars \citep{Kraus_2012}. Secondly, the intermediate stage of kilometer-sized planetesimal accretion is extremely sensitive to stellar companion perturbations \citep{Thebault_2006}. The formation of planets in close binaries or in the presence of a massive and inclined distant perturber \citep[inducing the Lidov-Kozai excitations of the protoplanetary disk; e.g.,][]{Martin_2014, Fu_2015_II, Zanazzi_2017a} is still a theoretical issue, since even moderate dynamical perturbations can generate high-velocity impacts and thus inhibit the formation of planetary embryos \citep{Lissauer_1993}. However, \cite{Batygin_2011} showed that disk self-gravity can be sufficient to suppress the Lidov-Kozai oscillations and maintain orbital coherence and planetary growth in the presence of a massive and inclined distant perturber. The decisive role that disk gravity plays in planetesimal dynamics was also confirmed by \cite{Rafikov_2015}. Lastly, it was shown that in  the late stage of planetary accretion, from Lunar-sized embryos to fully formed planets, the embryo accretion region roughly corresponds to the stability domain (e.g., \cite{Quintana_2007}). \cite{Haghighipour_2007} analyzed the formation of terrestrial planets in close binaries, assuming the existence of a giant planet already formed further out in the disk and a hundred planetary embryos (Moon- to Mars-sized bodies) formed in the region between the primary and the giant planet. These authors showed that planet formation can be efficient in moderately close binary planetary systems, and the final assembly of the planets and their water content strongly depend on the parameters of the stellar companion.

Regarding the formation of giant planet systems in binaries, much attention was directed toward the formation of hot Jupiters in binary star systems via Lidov-Kozai cycles and tidal friction \citep[e.g.,][]{Fabrycky_2007}. The planet-planet scattering mechanism was considered for Jupiter-mass planets orbiting the central star of a close binary system \citep{Marzari_2005}. Several hydrodynamical simulations of giant planets evolving in gas disks were performed for binary systems. Without providing an exhaustive list, we cite the following works. The evolution of planets embedded in circumbinary disks was considered for close binary stars, for instance, in \cite{Kley_2014} for  Kepler-38. \cite{Xiang_2014}, \cite{Picogna_2015}, and \cite{Lubow_2016} studied the evolution of a giant planet-disk system in the presence of an inclined binary companion and found that a substantial misalignment between the orbit of the planet and the disk generally occurs. \cite{Martin_2016} studied the formation of giant planets in Lidov-Kozai resonance with a highly misaligned binary companion, starting from a coplanar planet-disk system configuration. 

In the present work, we aim to realize a dynamical study of the influence of a wide binary companion on the evolution of a single giant planet migrating in the protoplanetary disk in the Type II regime. We particularly focus on how the Lidov-Kozai resonance articulates with the planetary migration when the stellar companion is highly inclined. Using a symplectic n-body code including eccentricity and inclination damping formulae to properly model the influence of the disk, we carry out 3200 numerical simulations with different initial eccentricity and inclination values of the binary companion. We follow the evolution of the giant planet when embedded in the protoplanetary disk and pursue the analysis well after the dispersion of the disk.

The paper is organized as follows. In Sect.~\ref{Simulations}, the code and initial settings of the simulations are described. In Sect.~\ref{Overall_results}, we present the results of the simulations by focusing on typical planetary evolutions as well as the final parameter distributions for the planet. In order to dynamically characterize the planetary evolutions, the results are discussed using a quadrupolar Hamiltonian approach in Sect.~\ref{Dynamical_analysis}. Finally, our conclusions are given in Sect.~\ref{Discussion}.

\section{Setup of the simulations}
\label{Simulations}

In this work, we focus on the evolution of a giant planet embedded in a protoplanetary disk undergoing Type II migration around one of the two stars of a wide binary star system (S-type planet). We denote with subscripts $0$ for the central star, $B$ for the binary companion, and $pl$ for the planet. The code used for the simulations and the initial parameters of our study are described in the following.

For our study we adapted the SyMBA n-body code \citep{Duncan_1998} to wide binaries by following the strategy of \citet{Chambers_2002}, based on a new set of coordinates (wide-binary coordinates). This last set induces a splitting of the Hamiltonian of the three-body problem composed of two stars and a planet, into three terms suitable for symplectic integration. Working in a Hamiltonian formalism, the family of symplectic integrators uses the symplectic structure of the problem to keep the Hamiltonian well conserved for a long-time scale (i.e., the energy error remains bounded), even when using a large time step. The evolution of the three terms can be performed separately via the Baker-Campbell-Hausdorff formula \citep{Bourbaki_1972}, upon which symplectic schemes such as that of \cite{Laskar_2001} have been built. However these schemes require that one part of the Hamiltonian is far larger than the other parts. The splitting that is classically used for the solar system (i.e., a term for the Keplerian motion and two terms depending on the position and momentum, respectively, for the mutual perturbation between the planets) is not suitable for binary systems because the perturbation is no longer smaller than the Keplerian part. The wide-binary coordinates of \citet{Chambers_2002} are defined such that all the large terms of the Hamiltonian can be incorporated into a single part of the Hamiltonian, making a symplectic integration possible with a leapfrog or $\rm{SABA}_1$ \citep{Laskar_2001} integrator.

\begin{table}
        \begin{tabular}{lrrr}
                \hline
                & Central star & Planet & Binary companion\\
                \hline
                mass  & $1$ $M_{\odot}$ & $U[1;5]$ $M_J$ & $1$ $M_{\odot}$\\
                $a$ (AU)  & & $15$ & $500$\\
                $e$  & & $U[0.001;0.01]$ & $10^{-5}$, $0.1$, $0.3$, $0.5$\\ 
                $i$ (°) & & $U[0.01;0.1]$ & $10^{-5}$, $10$, $20$, $30$, $40$,  \\
                & & & 50, 60, 70 \\
                $\Omega$ (°) & & $10^{-5}$& $10^{-5}$\\
                $\omega$ (°) & & $U[0;180]$& $10^{-5}$\\
                $M$ (°) & & $10^{-5}$ & $10^{-5}$\\
                \hline
        \end{tabular}
        \caption{Initial parameters of the simulations. The elements are associated with heliocentric coordinates with respect to the disk plane.}
        \label{Body_param}
\end{table}

The n-body code also includes the Type II migration of the giant planet caused by the angular momentum exchange with the protoplanetary gas disk \citep{Goldreich_1979}, mimicked by a suitable Stokes-type drag force added in the equations of motion. We used the same approach as in \cite{Sotiriadis_2017}, to which we refer for more details. The migration of a planet in Type II regime is on the same timescale as the viscous accretion time but, when the mass of the planet is comparable to the mass of the material in its vicinity, the migration rate scales with the ratio of the planetary mass over the local disk mass \citep{Ivanov_1999, Nelson_2000, Crida_2007}. Thus the timescale for Type II regime adopted in this work is written as
\begin{equation}\label{tmig}
\tau_{mig} = \frac{2}{3} \alpha^{-1} h^{-2} \Omega_{pl}^{-1}  \times max\left\{1, \frac{m_{pl}}{ (4 \pi/3) \Sigma(r_{pl}) r_{pl}^2}\right\},
\end{equation}
where $\alpha=0.005$  is the classical value for the Shakura-Sunyanev viscosity parameter \citep{Shakura_1973}, $h=0.05$ the disk aspect ratio, $\Omega_{pl}^{-1}=2\pi a_{pl}^{(-3/2)}$ the orbital frequency of the planet, $a_{pl}$ the semimajor axis of the planet, $m_{pl}$ the mass of the planet, and $\Sigma \propto r^{-0.5}$ the surface density profile of the disk. In the simulations, we fixed the disk inner and outer edges to $R_{in}=0.05$ AU and $R_{out}=30$ AU. The local disk mass is considered as the mass of the disk between $0.2 a_{pl}$ and $2.5a_{pl}$. The code also includes a smooth transition in the gas-free inner cavity using an hyperbolic tangent function $tanh \left(\frac{r-R_{in}}{\Delta r}\right),$ where $\Delta r=0.001$ AU, following \citet{Matsumoto_2012}.

Regarding the damping of the planetary eccentricity and inclination caused by the disk, we adopted the damping formulae of \citet{Bitsch_2013} achieved by three-dimensional hydrodynamical simulations of protoplanetary disks with embedded high-mass planets. The analytic formulae depend on the local mass disk, planetary mass, eccentricity, and inclination evaluated throughout the integration. These formulae are conceived for planets between 1 
$M_{J}$ and 10 $M_{J}$. We note that it is rather unclear if a highly inclined planet continues to migrate on a viscous accretion timescale. Nevertheless, we adopted this recipe since, when initially embedded in the disk, the planet only occasionally reached high inclination with respect to the disk plane during the protoplanetary disk phase (only a few simulations are concerned and only for a very brief period of time). No inclination damping is applied when $i_{pl} < 0.5^\circ$.

\begin{figure}[]
        \centering
        \resizebox{\hsize}{!}{\includegraphics{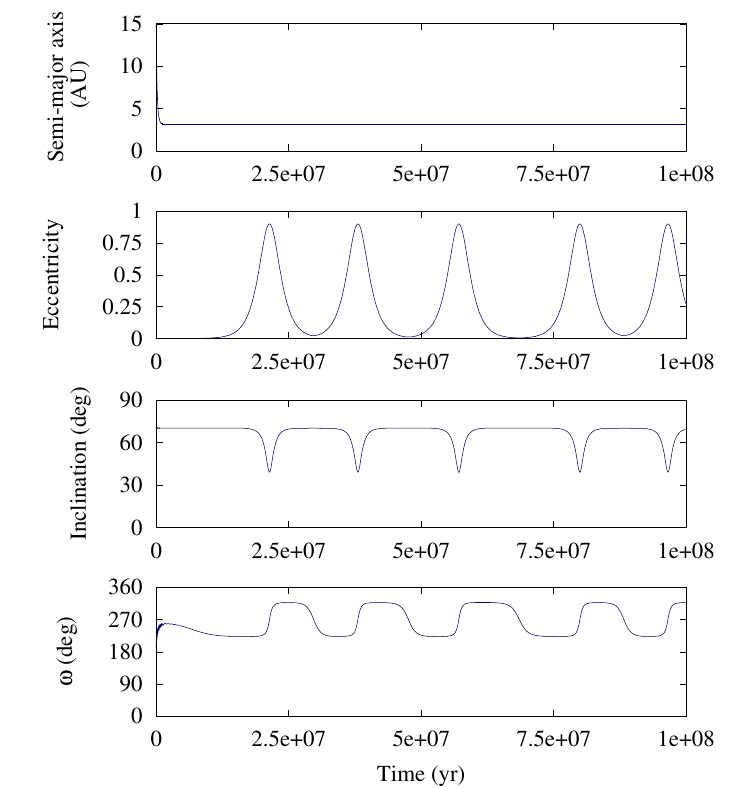}}
        \caption{Typical evolution of a Lidov-Kozai resonant system (libration of the pericenter argument in the invariant plane reference frame). The initial parameters are $m_{pl}=2.02 m_{Jup}$, $\omega_{pl}=16^\circ$, $e_B=0.3,$ and $i_B=70^\circ$.}
        \label{sys_2220}
\end{figure}

\begin{figure}[]
        \centering
        \resizebox{\hsize}{!}{\includegraphics{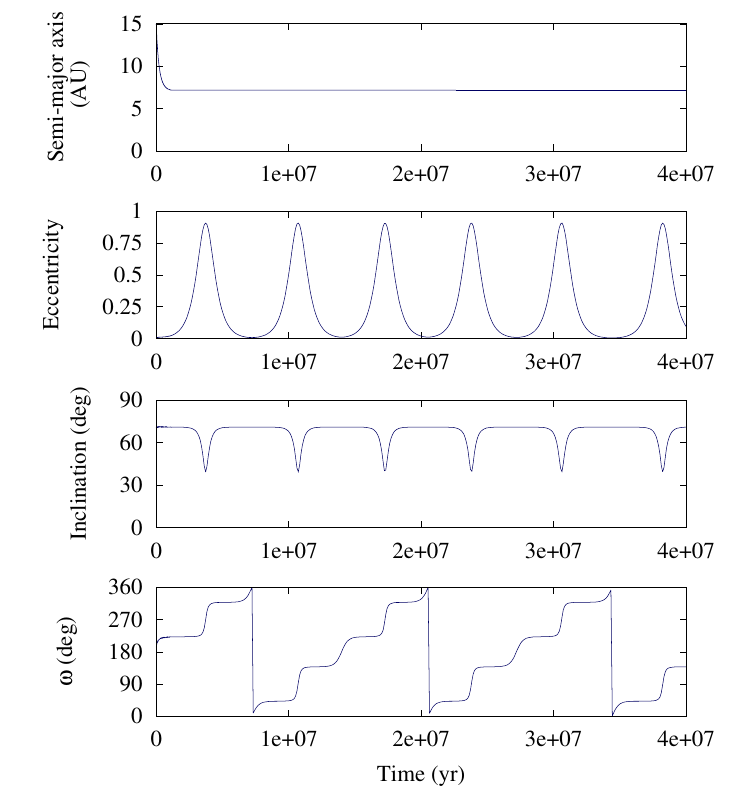}}
        \caption{Typical evolution of a planet that undergoes high eccentricity and inclination variations, while not in a Lidov-Kozai resonance (circulation of the pericenter argument in the invariant plane reference frame). The initial parameters are $m_{pl}=4.34 m_{Jup}$, $\omega_{pl}=13^\circ$, $e_B=10^{-5},$ and $i_B=70^\circ$.}
        \label{sys_479}
\end{figure}

Following \cite{Papaloizou_2000}, the orbital migration and the eccentricity and inclination damping effects are added to the acceleration of the planet as follows:
\begin{equation}
{\bf a_{disk}} = - \frac{{\bf v_{pl}}}{\tau_{mig}} -2\frac{({\bf v_{pl}} \cdotp {\bf r_{pl}}) \bf r_{pl}} {r_{pl}^2 \tau_{ecc}} -2 \frac{({\bf v_{pl}} \cdotp {\bf k}) {\bf k}}{\tau_{inc}},
\end{equation}
where ${\bf r_{pl}}$ is the position of the planet, ${\bf v_{pl}}$ the velocity of the planet, ${\bf k}$ the unit vector in the vertical direction, $\tau_{mig}$ the timescale for Type-II regime (see Eq. (\ref{tmig})), and $\tau_{ecc}$ and $\tau_{inc}$ the timescales for eccentricity and inclination damping (issued from \cite{Bitsch_2013}, see their formulae (11)-(16)), respectively. To keep the symmetry of the symplectic algorithm, the damping forces are implemented in the code as in \cite{Lee_2002}~as follows:
\begin{equation*}
f(t_{k+1}) = E_{disk}\left(\frac{\tau}{2}\right) \exp\left(T \tau \right)E_{disk}\left(\frac{\tau}{2}\right) f(t_{k}),
\end{equation*}
where $T$ is the linear operator associated to the Hamiltonian of the n-body problem and $E_{disk}\left(\cdot\right)$ stands for the evolution due to the disk. The time step of our simulations is fixed to  $0.001$ yr during the migration phase and $0.01$ yr after the dispersal of the disk, since the evolution of the system is then characterized by secular effects. The simulations are followed for 100~Myr.

The initial disk mass in our simulations is fixed to $8 M_{Jup}$ and decreases exponentially through the evolution of the system, with a dispersal time of $\sim 1$~Myr. More specifically, the disk is considered as dissipated and the interaction is neglected when 
\begin{equation*}
{d{m}_{disk}(t)}/{dt}={{m}_{disk}(0) \exp{(-t/T_0})}/{T_0}<10^{-9} m_0/yr,
\end{equation*}
where $m_0$ denotes the mass of the central star and $T_0$ $= 2.8\times 10^5$ yr. The self-gravity of the disk is not considered in this work.

The initial orbital elements and masses for the planet and the binary stars adopted in the simulations are summarized in Table~\ref{Body_param}. These are expressed in the disk plane reference frame. The wide binary companion has a mass of $1$ $M_{\odot}$ and is located at $500$ AU. The planetary mass $m_{pl}$, eccentricity $e_{pl}$, inclination $i_{pl}$, and argument of the pericenter $\omega_{pl}$ follow a uniform distribution. For the binary companion, we considered four values of the eccentricity $e_B$ (namely $10^{-5}, 0.1, 0.3$, and $0.5$) as well as eight values of the inclination $i_B$ (from $\sim0^\circ$ up to $70^\circ$). In total, by picking randomly 100 values for each uniform distribution, we carried out 3200 simulations. The dynamical evolutions observed in the simulations are presented in the next section.

\section{Results}
\label{Overall_results}

In this section, we describe in detail the long-term evolution of the planets as well as their orbital characteristics at the end of the simulations. Particular attention is given to the capture in the Lidov-Kozai resonance in the case of highly inclined binary companions.

\begin{figure*}[]
        \centering
        \includegraphics[width=0.49\linewidth]{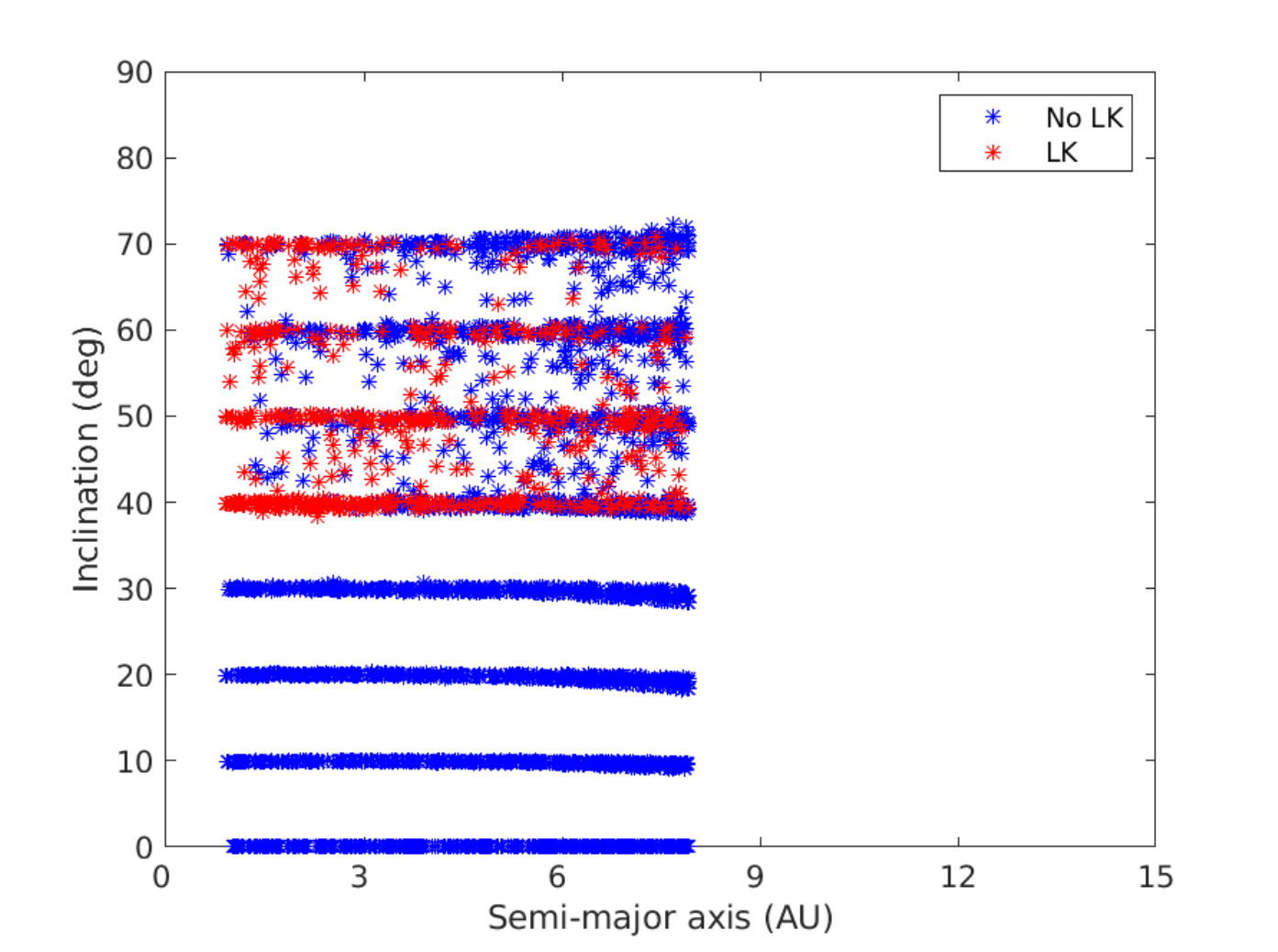}
        \includegraphics[width=0.49\linewidth]{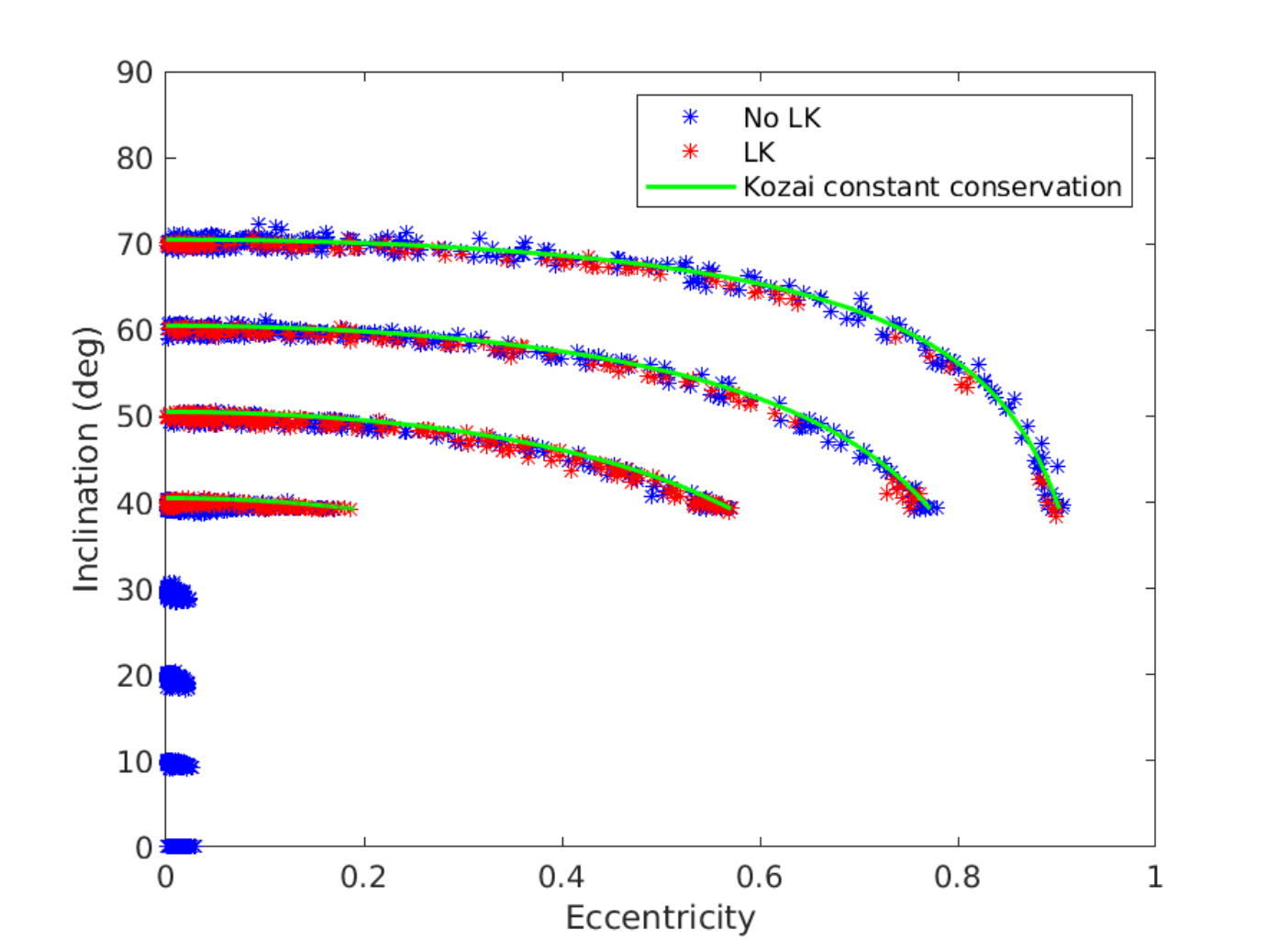}
        \caption{Inclination of the planets as a function of the semimajor axis (left panel) and the eccentricity (right panel) at the end of the simulations (100 Myr). The planets evolving in the Lidov-Kozai resonance are drawn in red.}
        \label{incl}
\end{figure*}

\begin{table}
                \centering
        \begin{tabular}{rrrr}
                $i_B$ ($^\circ$)  & Lidov-Kozai ($\%$)\\
                \hline
                $\le 30$ & 0 \\
                40 & 43  \\
                50 & 50  \\
                60 & 26  \\
                70 & 25  \\
        \end{tabular}
                \caption{Percentages of systems in a final Lidov-Kozai resonant state, for the different inclinations of the binary companion $i_B$.}
        \label{ejections}
\end{table}

\subsection{Lidov-Kozai resonance}\label{sec:LK}

In our simulations, a strong influence of the wide binary companion is observed on the evolution of the planet embedded in the protoplanetary disk, especially for highly inclined binary companions for which the Lidov-Kozai mechanism comes into play \citep{Lidov_1962,Kozai_1962}. When referred to the invariant plane orthogonal to the total angular momentum of the system (also called the Laplace plane), the planet in a Lidov-Kozai resonant state exhibits coupled variations in eccentricity and inclination as well as a libration of the pericenter argument around $\pm 90^\circ$. We note that the large orbital variations associated with the Lidov-Kozai resonance take place in a coherent way such that the long-term stability of the system is assured. A typical example of a Lidov-Kozai evolution is shown in Fig.~\ref{sys_2220}, for an inclination $i_B=70^\circ$ of the binary companion with respect to the disk plane. This inclination value is higher than the critical value $\arccos{\sqrt{3/5}}\approx 39.23^\circ$ introduced by \cite{Kozai_1962} for the restricted three-body problem. After the migrating phase, the planet is rapidly captured in a Lidov-Kozai resonance, as indicated by the libration of the pericenter argument observed in the last panel of Fig. 1, and high eccentricity and inclination variations are observed.

It is important to note that the libration of the pericenter argument in the invariant plane reference frame is the proper criteria for the identification of the Lidov-Kozai resonant regime among the simulations \citep{Libert_2009}. High eccentricity and inclination variations can also be observed when the planet is not in a Lidov-Kozai resonant state, as shown in Fig.~\ref{sys_479}. While the planet does experience large variations in eccentricity and inclination, the argument of the pericenter circulates (see the last panel of Fig. 2). For this reason, all the dynamical evolutions displayed in the following are presented in the invariant plane reference frame. 

Note that, in practice, a planet will be considered here as locked in the Lidov-Kozai resonance if a libration of the argument of the pericenter is observed at the end of the simulation during three Lidov-Kozai cycles (for the calculation of the period, see e.g. \cite{Innanen_1997}) or, in case it exceeds the timescale of the simulation, during the last $8\times10^7$~yr.

\begin{figure*}[]
        \centering
        \includegraphics[width=0.49\linewidth]{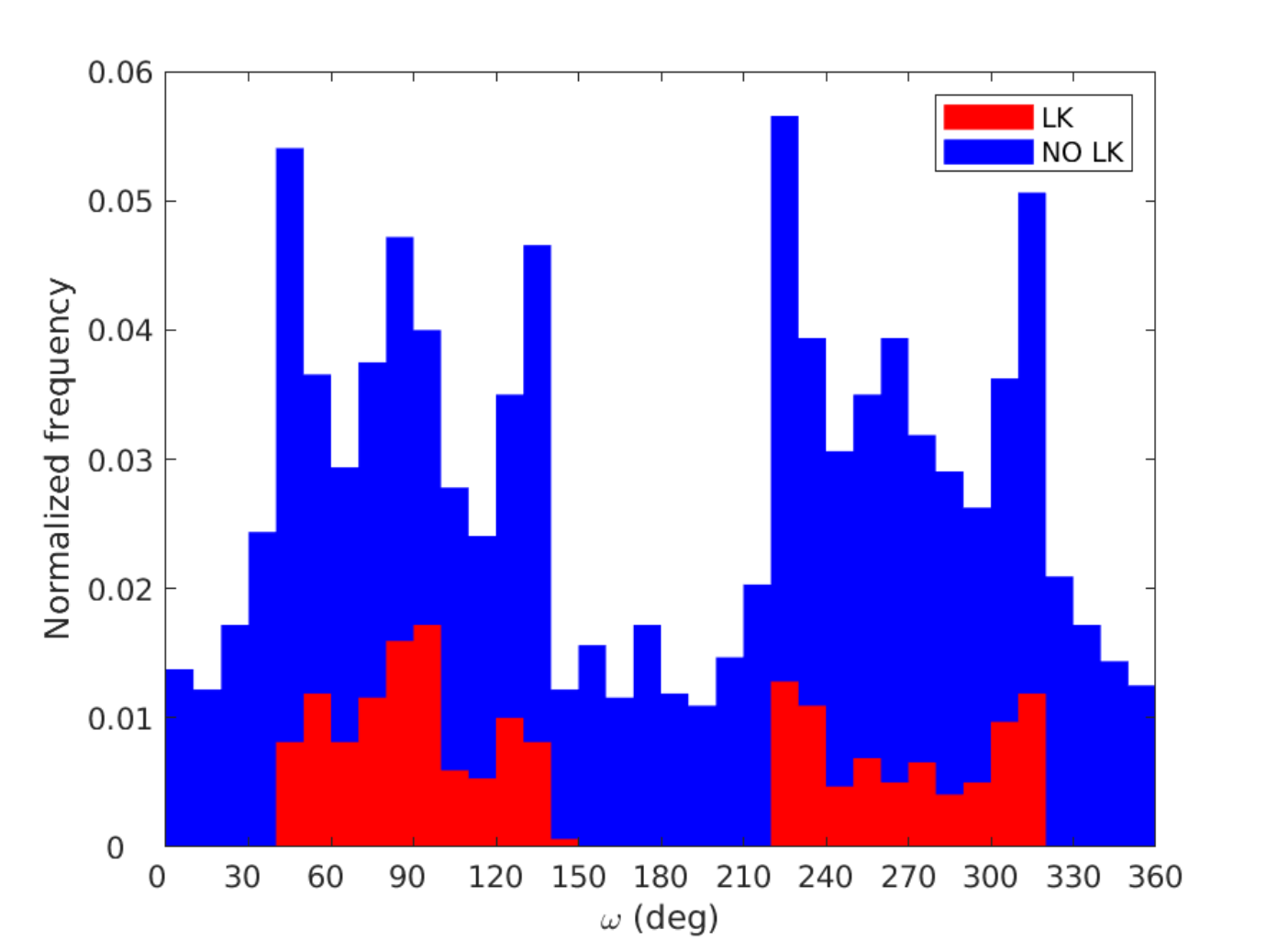}
        \includegraphics[width=0.49\linewidth]{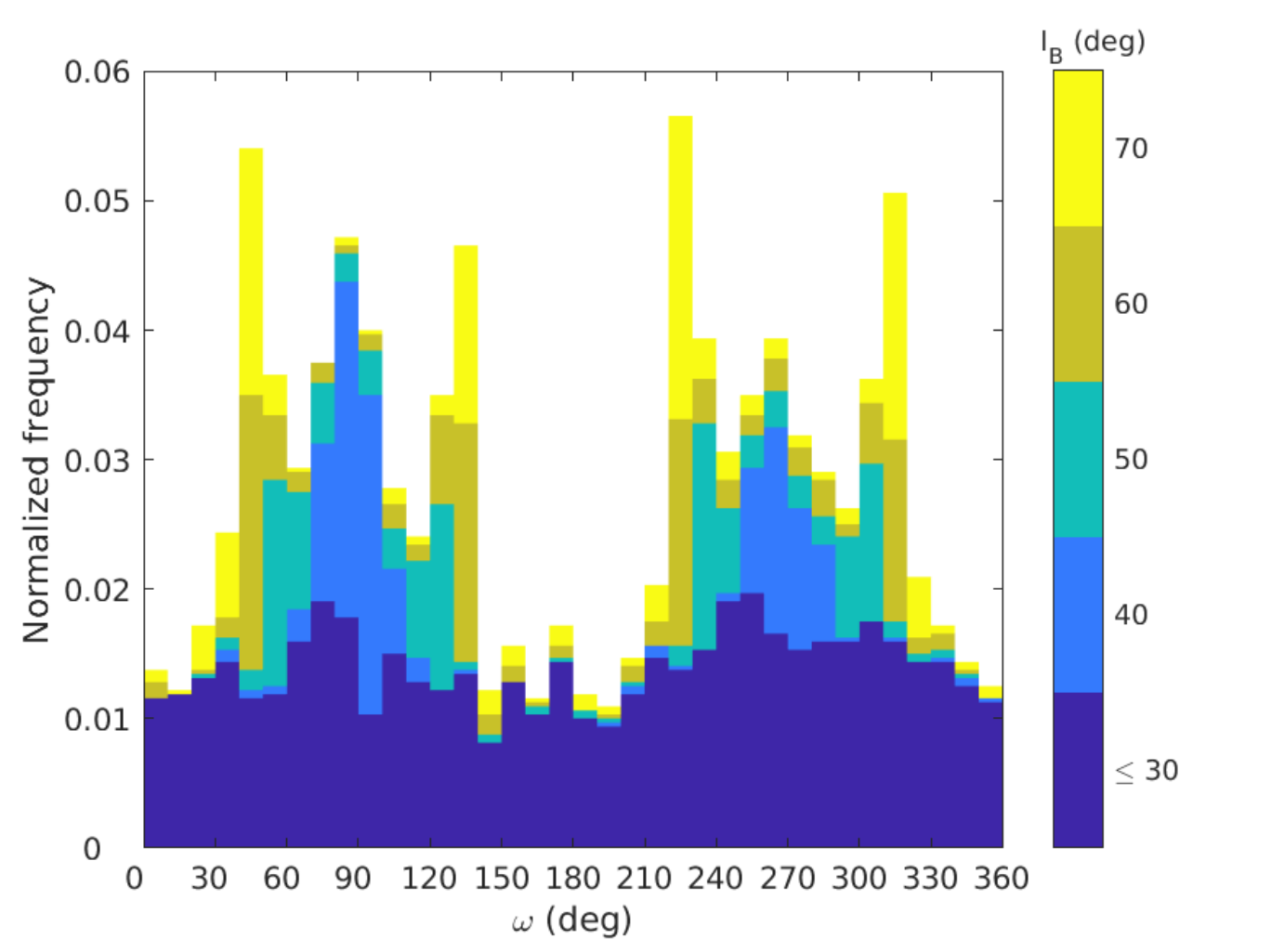}

        \caption{Normalized distribution of the pericenter argument of the planet, at the end of the simulation (100 Myr). The color codes refer to planets evolving in the Lidov-Kozai resonance in the left panel (red color) and to the initial inclination of the binary companion $i_B$ in the right panel.}
        \label{Hist_w}
\end{figure*}

\begin{figure*}[]
        \centering
        \includegraphics[width=0.49\linewidth]{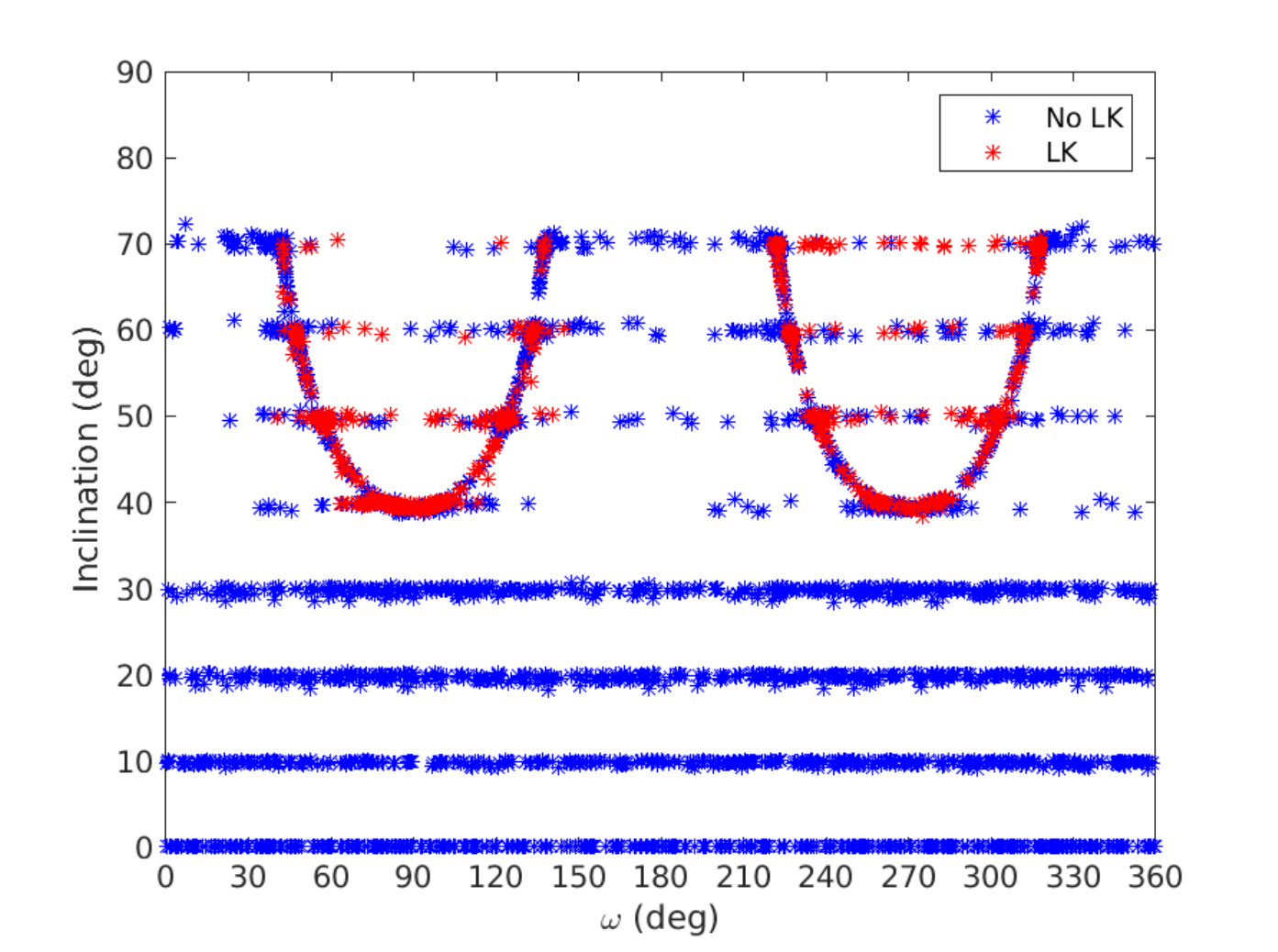}
        \includegraphics[width=0.49\linewidth]{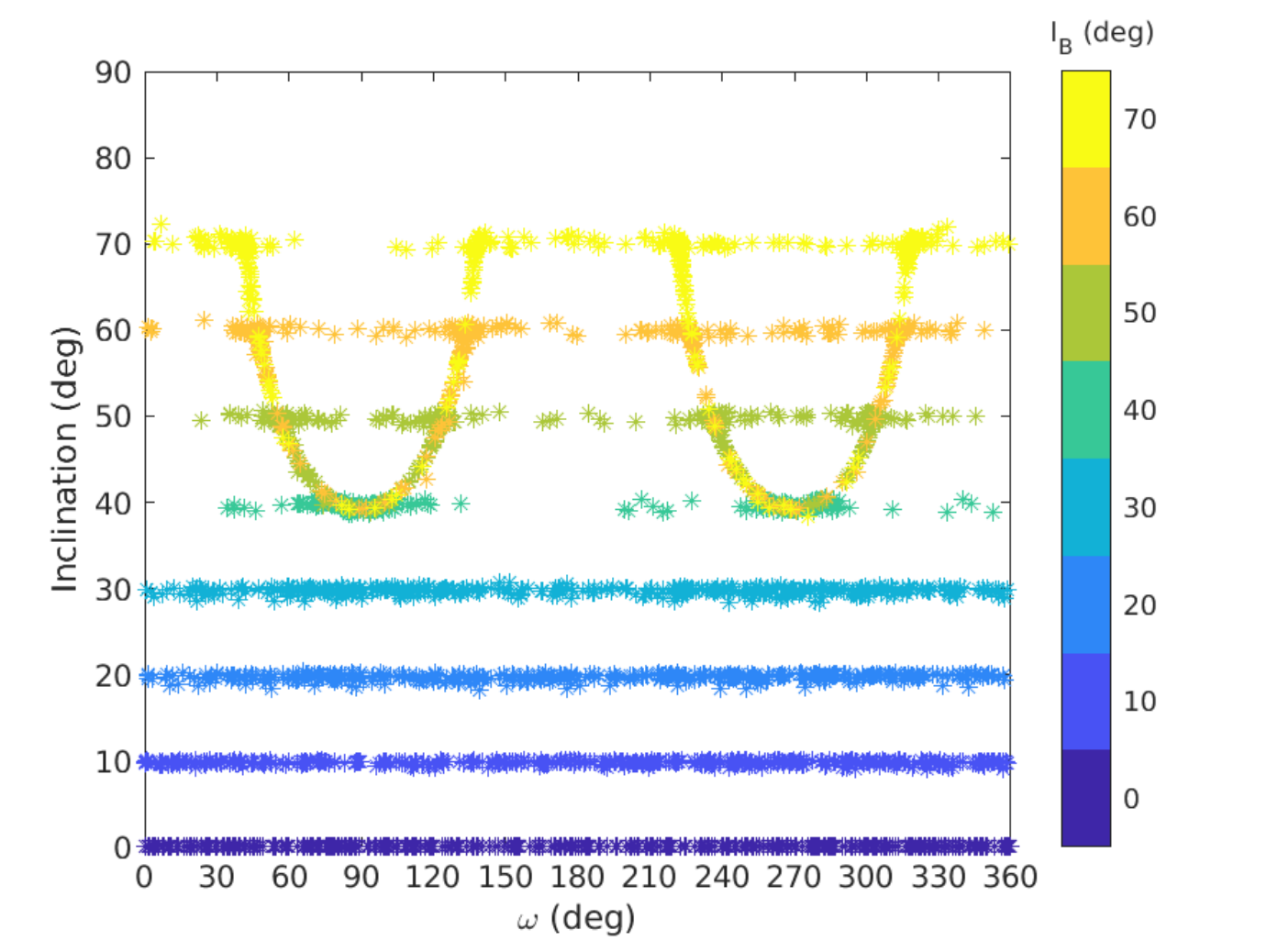}
        \caption{Inclination of the planet as a function of the argument of the pericenter, at the end of the simulation (100 Myr). The color codes refer to planets evolving in the Lidov-Kozai resonance in the left panel (red color) and to the initial inclination of the binary companion $i_B$ in the right panel.}
        \label{i_w}
\end{figure*}

\begin{figure}[]
        \centering
        \subfloat{\resizebox{0.95\hsize}{!}{\includegraphics{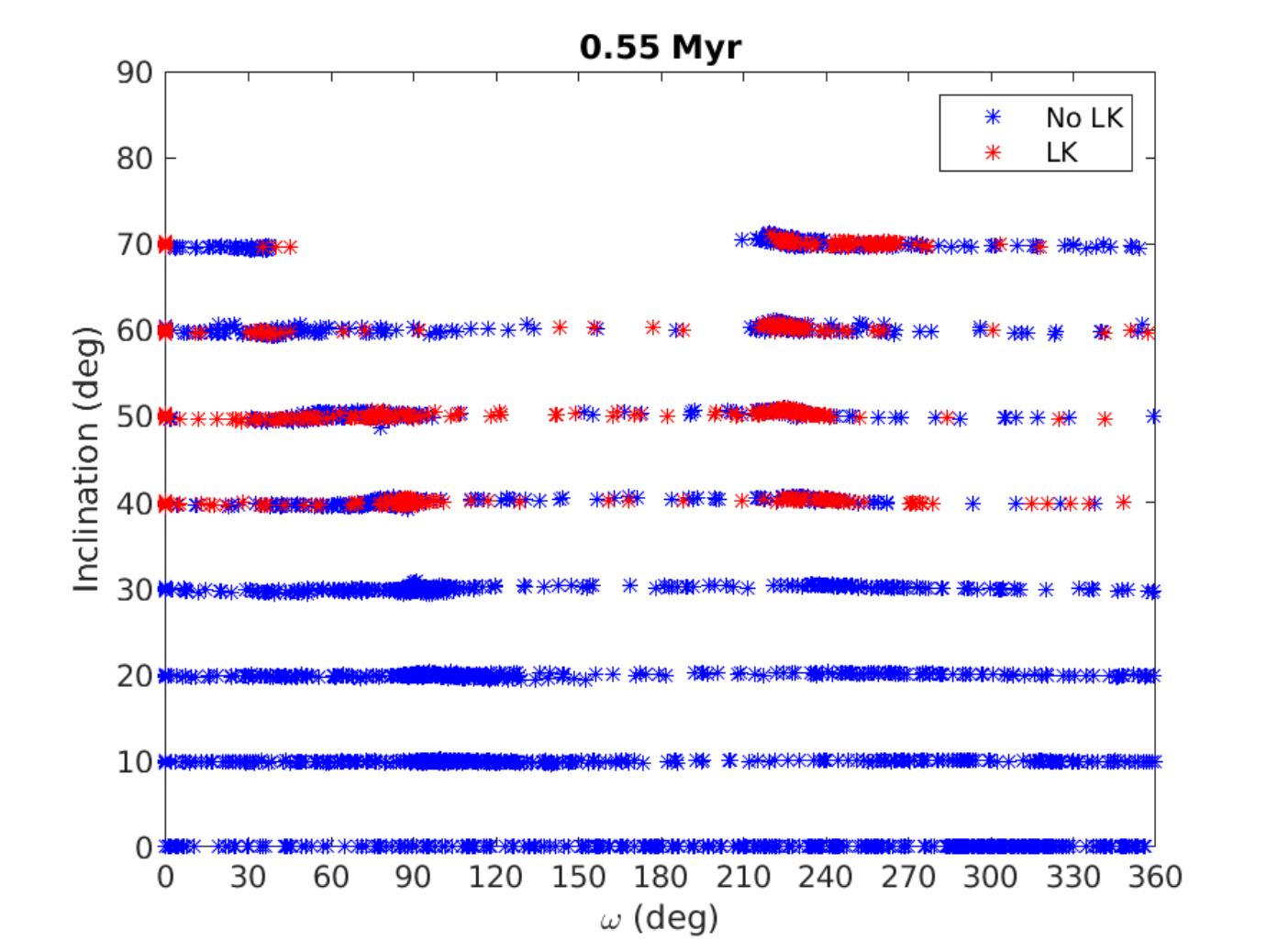}}}\\
        \subfloat{\resizebox{0.95\hsize}{!}{\includegraphics{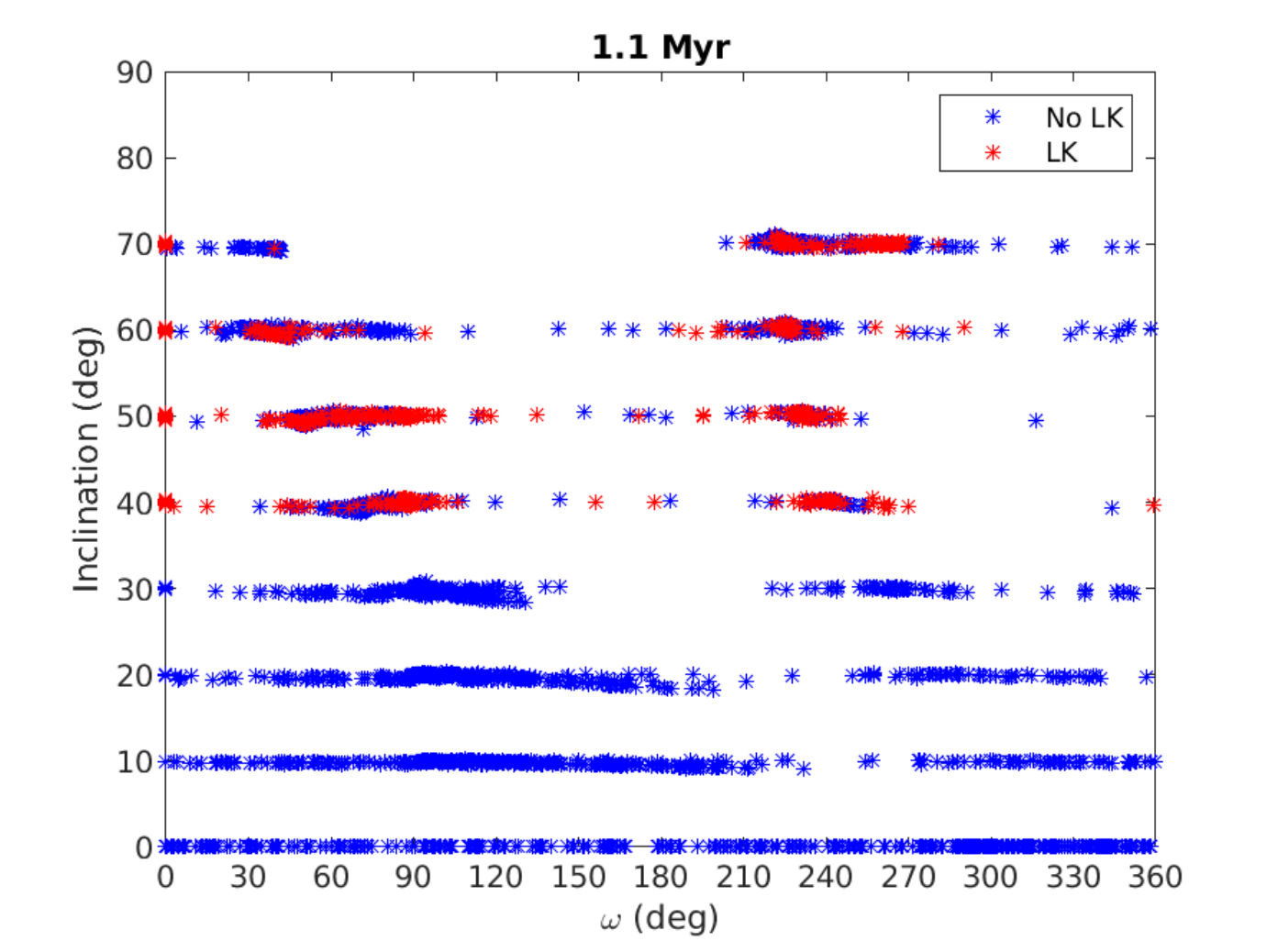}}}\\
        \subfloat{\resizebox{0.95\hsize}{!}{\includegraphics{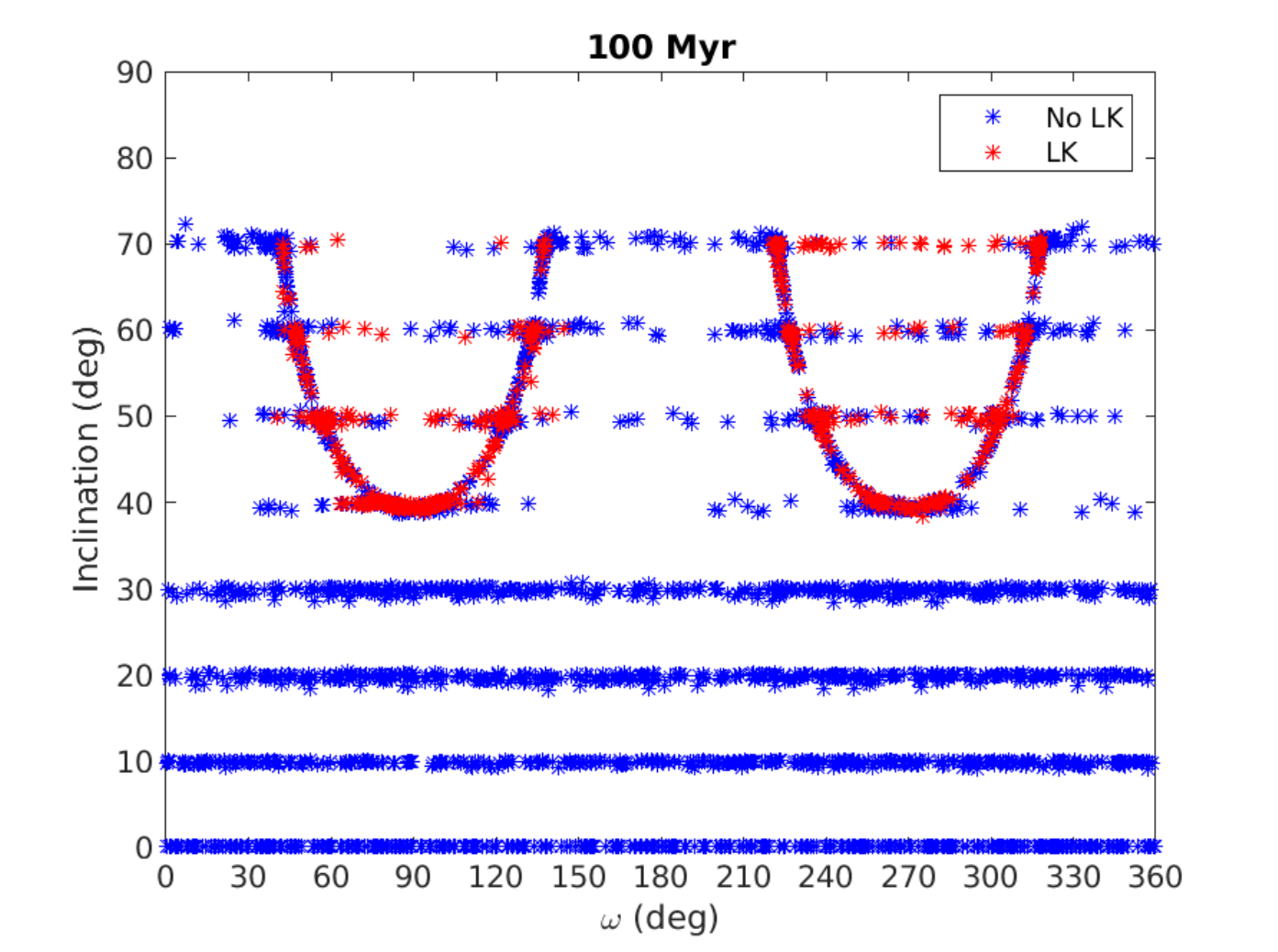}}}
        \caption{Same as Fig.~\ref{i_w}, for different times: $0.55$ Myr (half of the disk lifetime) in the top panel, $1.1$ Myr (dispersion of the disk) in the middle panel, and $100$ Myr (end of the simulation) in the bottom panel.}
        \label{evolution_w}
\end{figure}

\subsection{Final distributions}\label{sec:final}
In this section, we describe the orbital parameters of the planets at the end of the simulations (i.e., 100 Myr). We recall that all the planetary orbital parameters are given in the invariant plane reference frame. Fig.~\ref{incl} shows the planetary inclination as a function of the semimajor axis and the eccentricity. The color code indicates if the planet is in a Lidov-Kozai resonance state at the end of the evolution.

As seen in the left panel of Fig.~\ref{incl}, all the planets end up with a semimajor axis that is smaller than 10 AU as a result of the Type II migration of the planets during the protoplanetary disk phase. The planets found locked in a Lidov-Kozai resonant state (red stars) all have an inclination (with respect to the invariant plane) that is higher than $\sim 40^\circ$, which comes as no surprise considering the critical value of $39.23^\circ$ for the establishment of the Lidov-Kozai resonance. Nevertheless, the highly inclined planets are not all in a Lidov-Kozai resonant state at the end of the simulations, as shown by the color code.

The critical inclination value is even more obvious in the right panel of Fig.~\ref{incl}, where the planetary inclinations are presented jointly with the eccentricities. While a pileup around low eccentricities is observed at low planetary inclinations, high eccentricities are only associated with inclinations above $\sim 40^\circ$. However, although the Lidov-Kozai resonance could contribute to the excitation of the planetary eccentricities, we note that only a small fraction of the planets evolves in a Lidov-Kozai resonance at the end of the simulations. The percentages of final Lidov-Kozai resonant evolutions found among the 3200 simulations are given in Table~\ref{ejections}, for each initial inclination value of the binary companion. Let us remark that the percentages of resonance captures are slightly lower when the simulations do not include the disc phase. It seems to indicate that the migration in the disc promotes the establishment of Lidov-Kozai resonant sytems. Additionally, we note that the values in eccentricity and inclination displayed in Fig.~\ref{incl} follow particular curves (green lines) whose trend will be discussed in Section~\ref{Dynamical_analysis}.

Since the argument of the pericenter plays a key role in the Lidov-Kozai mechanism, we therefore pay particular attention to this angle in the following. Fig.~\ref{Hist_w} shows the normalized distribution of the argument of the pericenter of the planets at the end of the simulations. We clearly observe two pileups around $90^\circ$ and $270^\circ$. Interestingly, the highest peaks are found at the borders of the two pileups, namely for pericenter arguments close to $45^\circ$, $135^\circ$, $225^\circ$, and $315^\circ$. In the left panel of Fig.~\ref{Hist_w}, the systems in a Lidov-Kozai resonance belong to the pileups at $90^\circ$ and $270^\circ$, as expected. In the right panel of Fig.~\ref{Hist_w}, the color code of the histogram refers to the initial inclination values of the binary companion $i_B$. We can deduce from the plot that the pileups are formed by the highly inclined binary companions ($i_B \ge 40^\circ$), since the normalized frequencies of the pericenter arguments for planets with a binary companion initially below $40^\circ$  are almost uniformly distributed.

Finally, in Fig.~\ref{i_w}, we present the inclination of the planet as a function of the argument of the pericenter, at the end of the simulation, with the same color codes as in Fig.~\ref{Hist_w}. We note the existence of two arc-shaped curves. The Lidov-Kozai resonant systems mainly lie around these two curves, as observed in the left panel. In the right panel, we see that the two arc-shaped curves mostly correspond to planets with highly inclined binary companions. We note that the accumulations observed at the borders of the two pileups in the histograms of Fig.~\ref{Hist_w} can be connected with the particular shape of these curves. The edges of the arc-shaped curves are nearly parallel to the $y$-axis and thus more systems are gathered at the edges rather than in the central part of the curves (i.e., around $90^\circ$ and $270^\circ$) when considering parts of the $x$-axis of the same length.

\subsection{Initial conditions leading to Lidov-Kozai resonance}\label{sec:CI}

We might wonder whether it is possible to identify the initial planetary orbital elements leading to a Lidov-Kozai resonant state for the planet. As can be deduced from Table~\ref{ejections}, approximately 36$\%$ of the planets with a binary companion inclined at more than $40^\circ$ end up trapped in a Lidov-Kozai resonance at the end of the simulation. A statistical analysis of the initial parameters of these systems was performed, but no trend was observed. In particular, the initial values of the pericenter argument of the planets captured in the Lidov-Kozai resonance follow an uniform distribution. This is further illustrated in the evolutions of Fig.~\ref{sys_2220} and Fig.~\ref{sys_479}, which have close initial values of the arguments of pericenter but different resonant behaviors.

Moreover, we carried out  a study of the specific time of the capture in the Lidov-Kozai resonance. We present in Fig.~\ref{evolution_w} the same plot as in the left panel of Fig.~\ref{i_w}, but for different times. The top panel shows the planetary inclination versus the pericenter argument values at half of the disk lifetime ($0.55$ Myr), the middle  panel at the dispersion of the disk ($1.1$ Myr) and the bottom panel at the end of the simulation ($100$ Myr). We clearly see the accumulation of the planets around the arc-shaped curves through time. Interestingly, the planets firstly gather in the left part of the curves (i.e., around a pericenter argument of $45^\circ$ and $135^\circ$) during the disk phase and spread along the curves in a second phase.

\begin{figure*}[]
        \centering
        \includegraphics[width=0.4\linewidth]{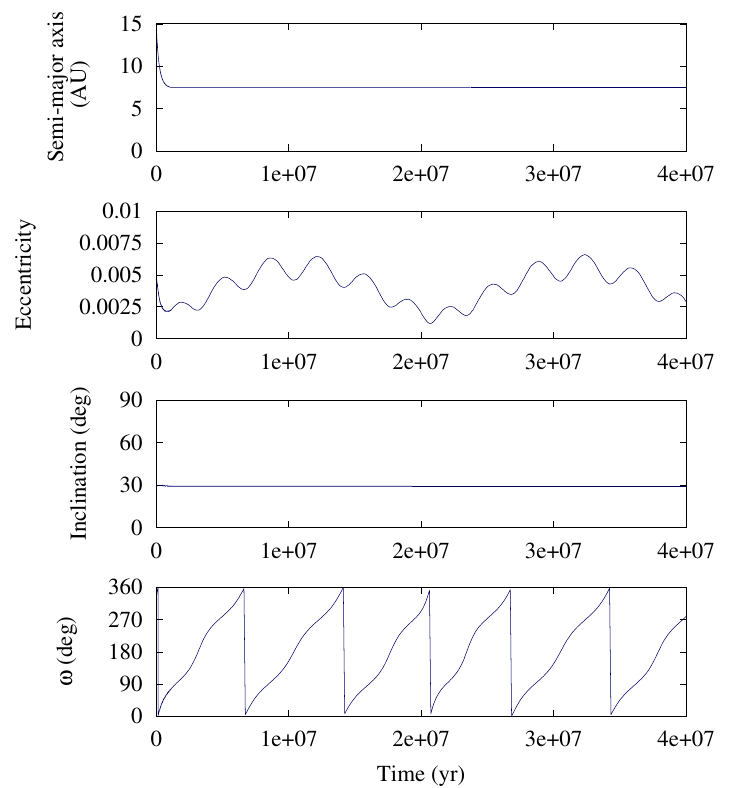} \hspace{0.5cm}
        \includegraphics[width=0.55\linewidth]{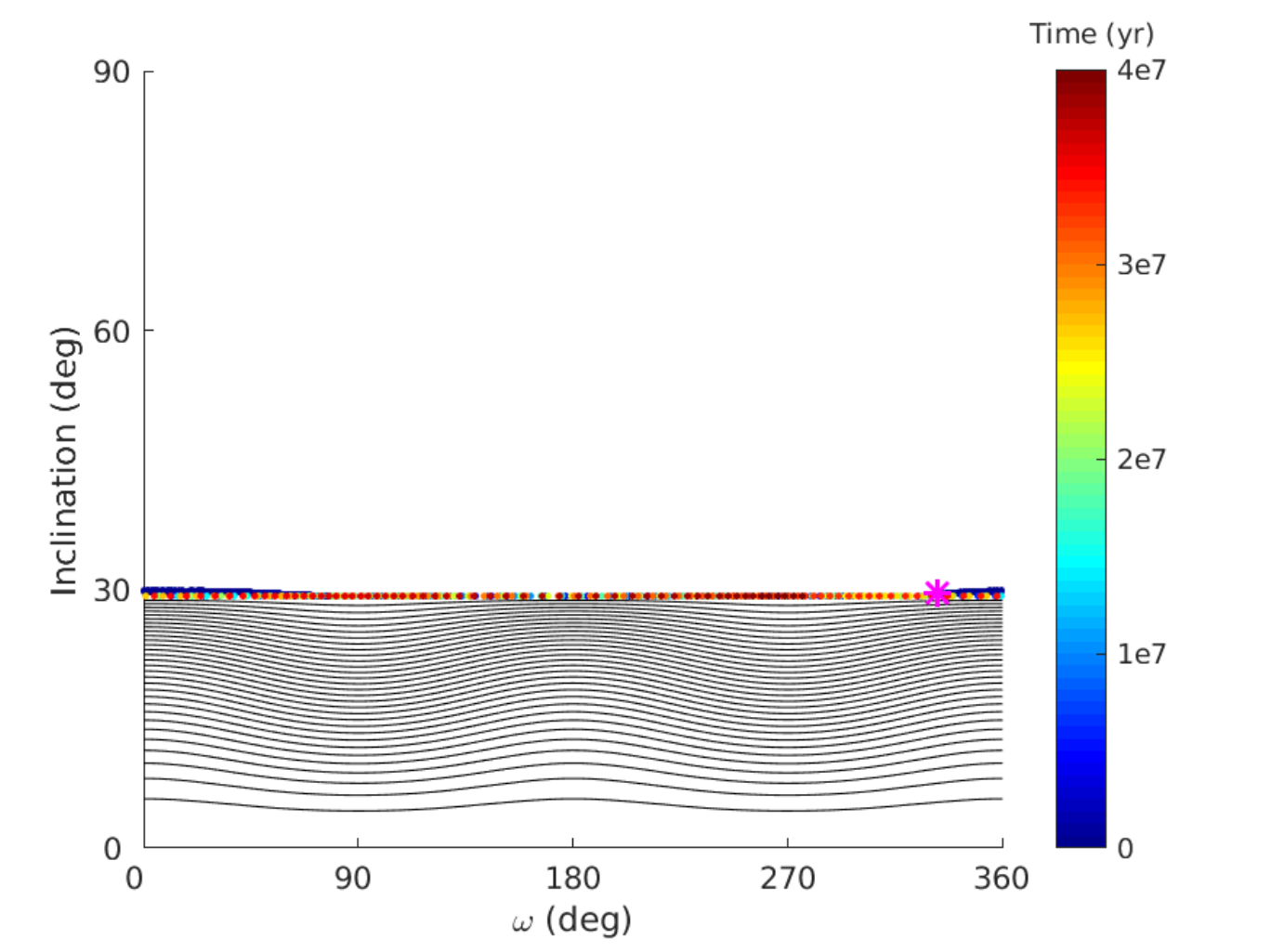}
        \caption{Left panel: Typical evolution of a planet in the presence of a binary companion with an inclination lower than $40^\circ$. Right panel: Phase portrait of the evolution. The initial parameters are $m_{pl}=4.6~m_{Jup}$, $\omega_{pl}=152^\circ$, $e_B=0.1,$ and $i_B=30^\circ$.}
        \label{sys_1168}
\end{figure*}

\begin{figure*}[]
        \centering
        \includegraphics[width=0.49\linewidth]{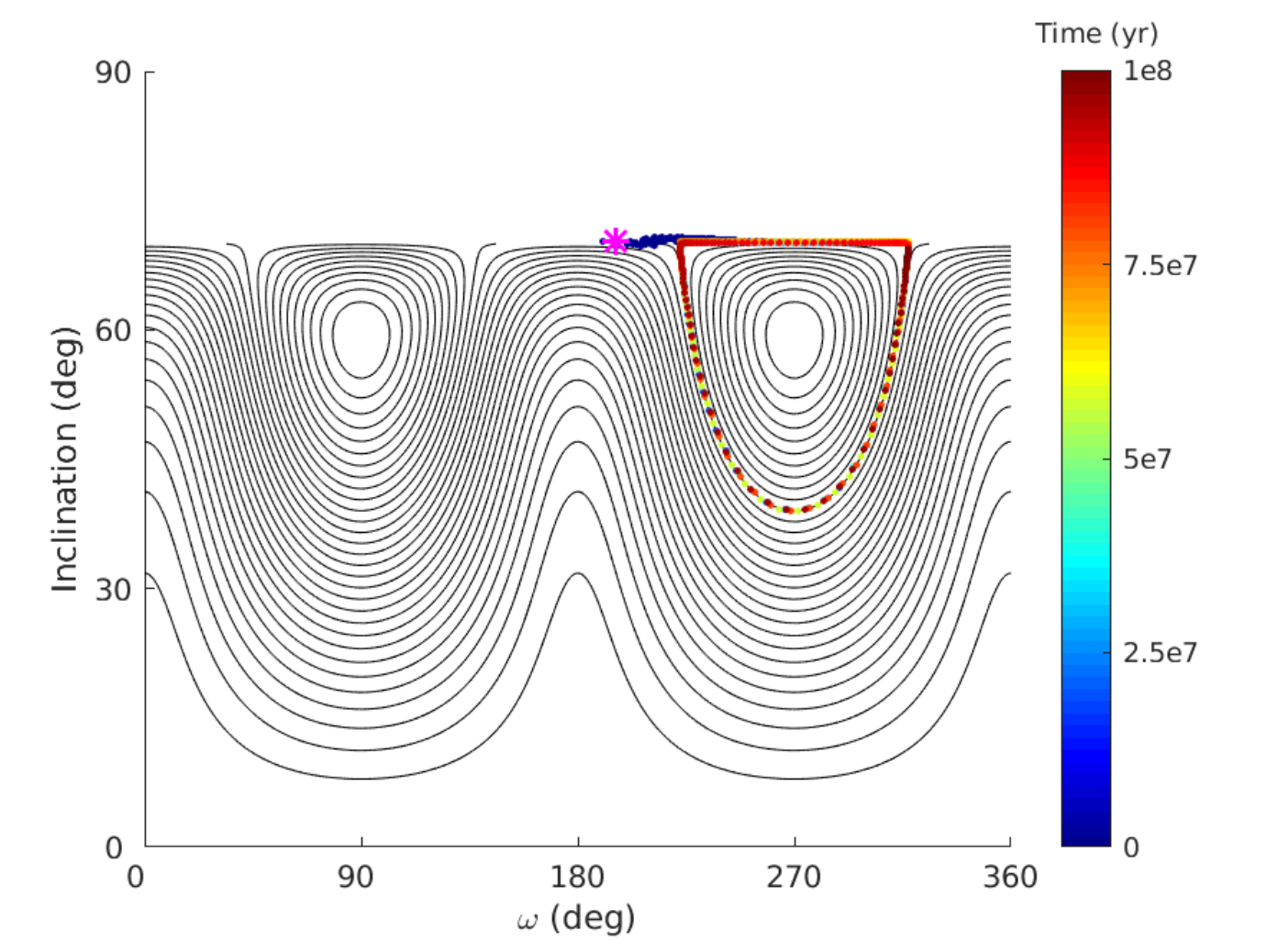}
        \includegraphics[width=0.49\linewidth]{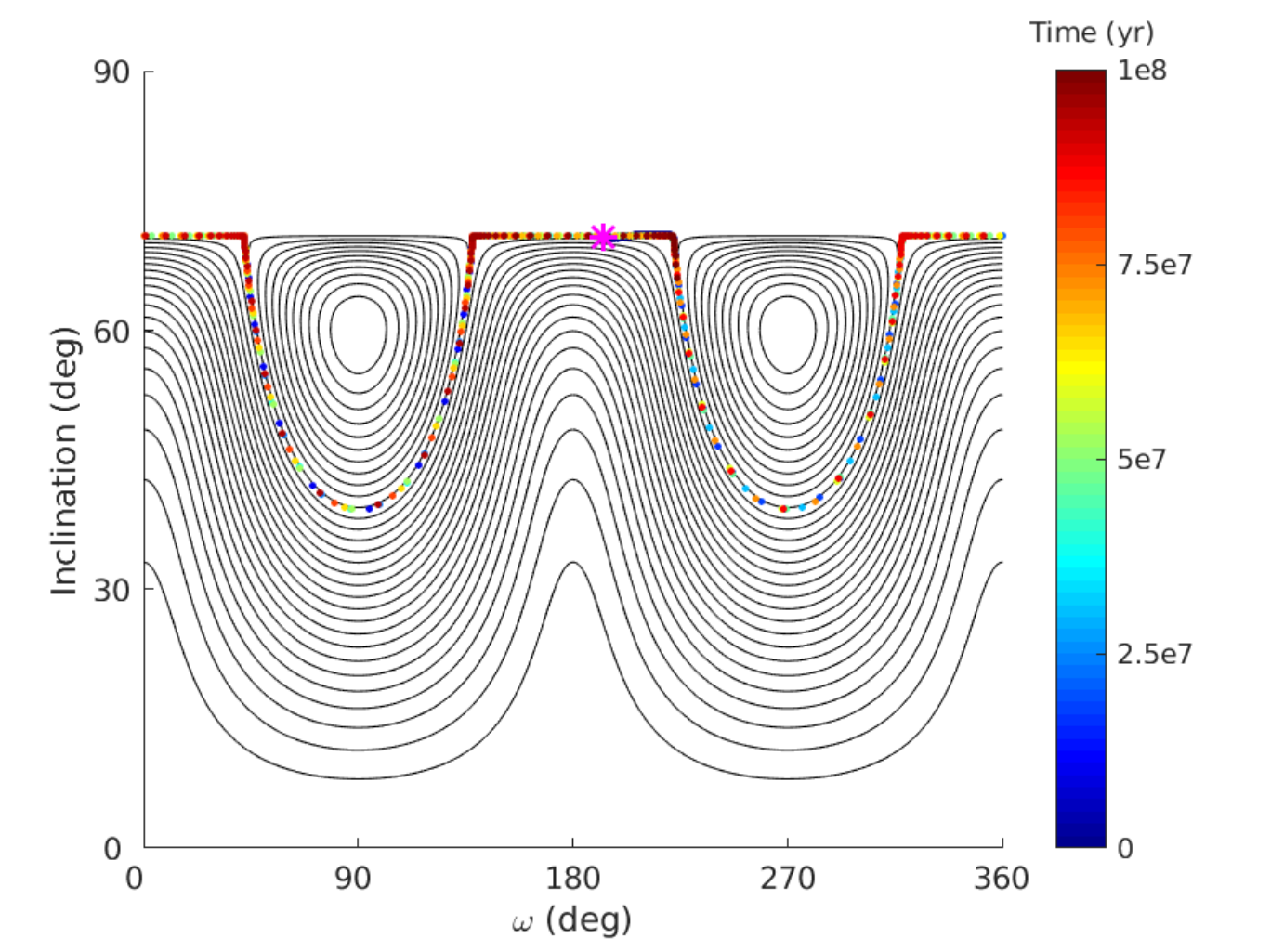}
        \caption{Phase portraits of different planetary evolutions in the presence of a binary companion at $i_B=$70$^{\circ}$. Left: The Lidov-Kozai resonant evolution from Fig.~\ref{sys_2220}. Right: The non Lidov-Kozai resonant evolution from Fig.~\ref{sys_479}.}
        \label{ana_kozai}       
\end{figure*}

\begin{figure*}[]
	\centering
	\includegraphics[width=0.4\linewidth]{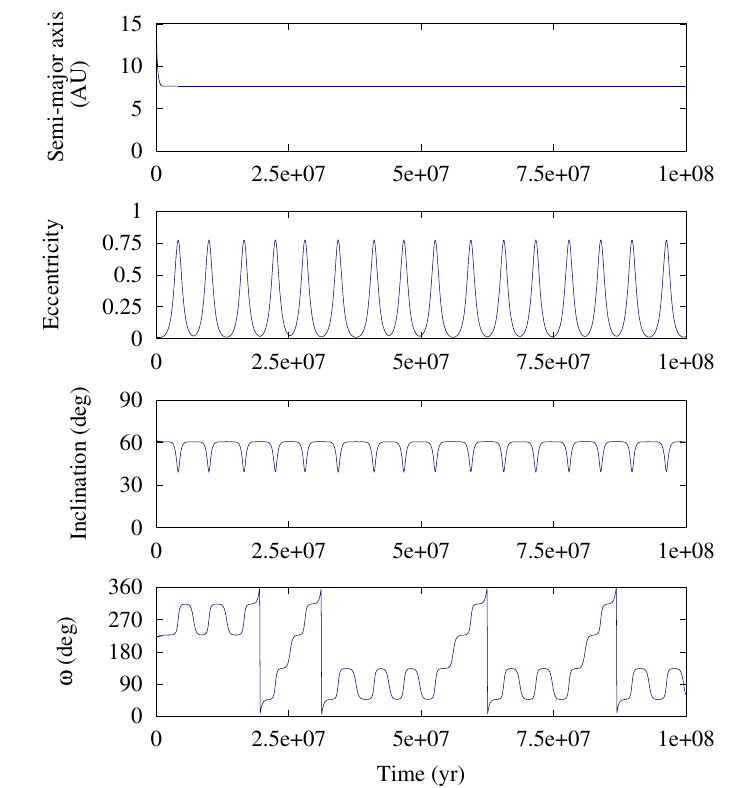} \hspace{0.5cm}
	\includegraphics[width=0.55\linewidth]{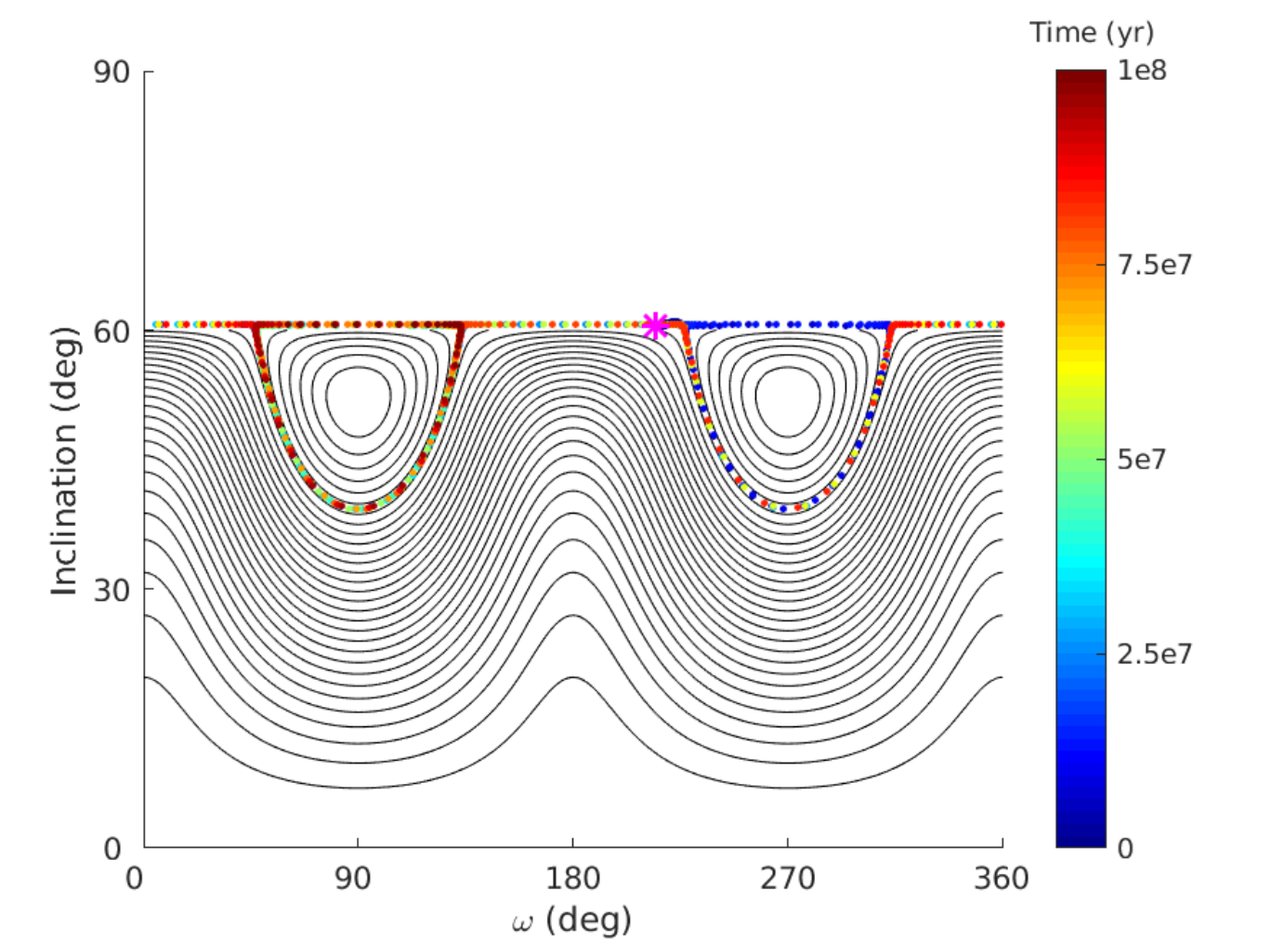}
	\caption{Left panel: Typical evolution of a planet alternating between resonance capture and circulation. Right panel: Phase portrait of the evolution. The initial parameters are $m_{pl}=4.74~m_{Jup}$, $\omega_{pl}=35^\circ$, $e_B=0.1,$ and $i_B=60^\circ$.}
	\label{sys_2618}
\end{figure*}

\section{Dynamical analysis}
\label{Dynamical_analysis}

In this section, we give a dynamical interpretation of the arc-shaped curves observed in Fig.~\ref{i_w}. To describe the dynamics of the migrating planets, we draw phase portraits of a simplified Hamiltonian formulation and follow the trajectory of the planets on this portrait until their final location.

Following the work of \cite{Bataille_2018}, we used a quadrupolar Hamiltonian approach (i.e., the dominating terms of the secular Hamiltonian) that has one degree of freedom and states as follows: 
\begin{multline*}
\mathcal{H} = -\left(5-3\frac{h^2}{\cos^2i_{pl}}\right)(1-3\cos^2(i_{pl}))\\ +15\left(1-\frac{h^2}{\cos^2i_{pl}}\right)(1-\cos^2i_{pl}) \cos(2\omega_{pl}).
\end{multline*}
Since the angular momentum of the wide binary orbit is much greater than that of the planetary orbit, the dimensionless angular momentum of the planet $h=\sqrt{1-e_{pl}^2}\cos i_{pl}$ is nearly constant and is usually referred to as the Kozai constant. For a given value of the Kozai constant, phase portraits of the dynamics can be drawn, consisting of level curves of the Hamiltonian in the plane $(\omega_{pl}, e_{pl})$ (or equivalently $(\omega_{pl}, i_{pl})$). 

For a low value of the inclination of the binary companion, the Hamiltonian curves in the plane $(\omega_{pl}, e_{pl})$ are nearly straight lines. An example of such an evolution is shown in Fig.~\ref{sys_1168}, for an inclination of $30^\circ$ for the companion star. In the left panel, we see that no significant eccentricity and inclination variation is observed during the long-term evolution of the planet. This evolution is overplotted on the phase portrait given in the right panel of Fig.~\ref{sys_1168}. The time evolution is indicated by the color scale, while the initial planetary parameters are denoted with a magenta star symbol. The value of the Kozai constant $h$ used to compute the phase portrait is fixed to the mean value of $h$ after the migration phase. Thus, at the beginning (dark blue points), the planetary evolution can depart from the Hamiltonian curves represented on the phase portrait (since the system is dissipative owing to the migration), but after the migration phase the trajectory of the planet evolves on a fixed level curve of the analytic model. In Fig.~\ref{sys_1168}, the planet follows a quasi-straight line of constant Hamiltonian, as expected.

For systems with a highly inclined companion ($i_B \ge$ 40°), the phase space shows the Lidov-Kozai resonance islands that strongly influence the dynamical behavior of the planet. Two typical dynamical outcomes are observed in the simulations and these are both illustrated in Fig.~\ref{ana_kozai}. In the left panel, we show the phase portrait for the system of Fig.~\ref{sys_2220}, for which a capture around the Lidov-Kozai equilibrium around $90\degree$ takes place. The evolution of the system of Fig.~\ref{sys_479} is represented on the phase space of the right panel. No capture in the Lidov-Kozai resonance is observed, and the planet evolves around the two islands with its pericenter argument circulating. Finally, we note that in both cases the planets spend time in libration or circulation around the Kozai equilibria. This gives an explanation for the arc-shaped curves observed in the final distribution of the simulations in Fig.~\ref{i_w}. 

Having a careful look to the two phase spaces of Fig.~\ref{ana_kozai}, one could expect that the points along the horizontal lines inside the two arc-shaped curves in Fig.~\ref{i_w} are associated with Lidov-Kozai resonant motion (red color). The blue points on the horizontal lines correspond to planets that alternate between phases of Lidov-Kozai resonant captures and phases of circulation around the islands. An example of such an evolution is represented on Fig.~\ref{sys_2618}. The planet is first captured around the island centred on $\omega = 90^\circ$ at the dispersion of the disk, then after some cycles leaves the resonance to circulate around the islands, before being captured around the second island centred on $\omega = 270^\circ$ and starting to alternate between circulation and resonance capture.

\begin{figure*}[]
        \centering
        \subfloat{\resizebox{0.33\hsize}{!}{\includegraphics{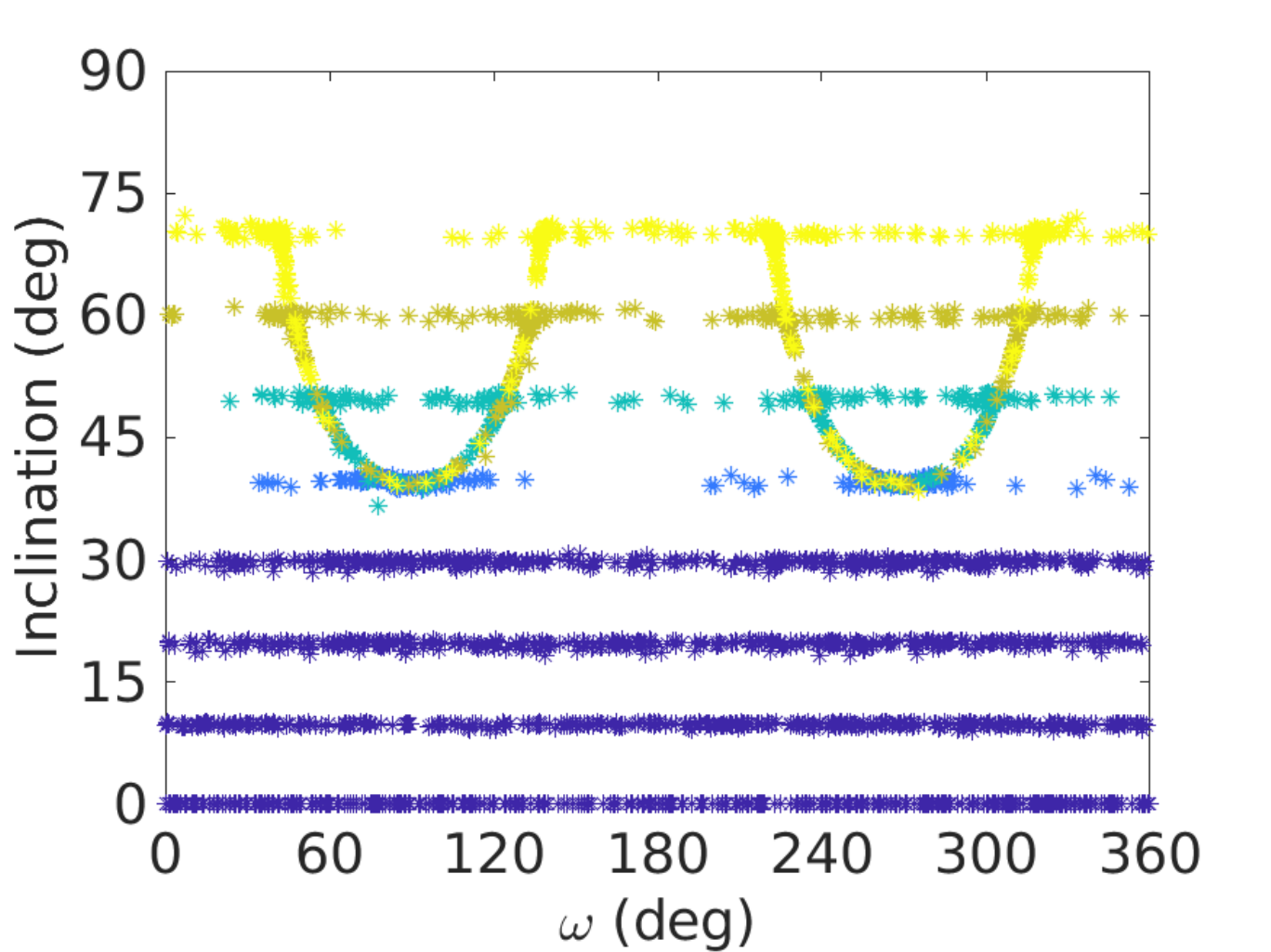}}}
        \subfloat{\resizebox{0.33\hsize}{!}{\includegraphics{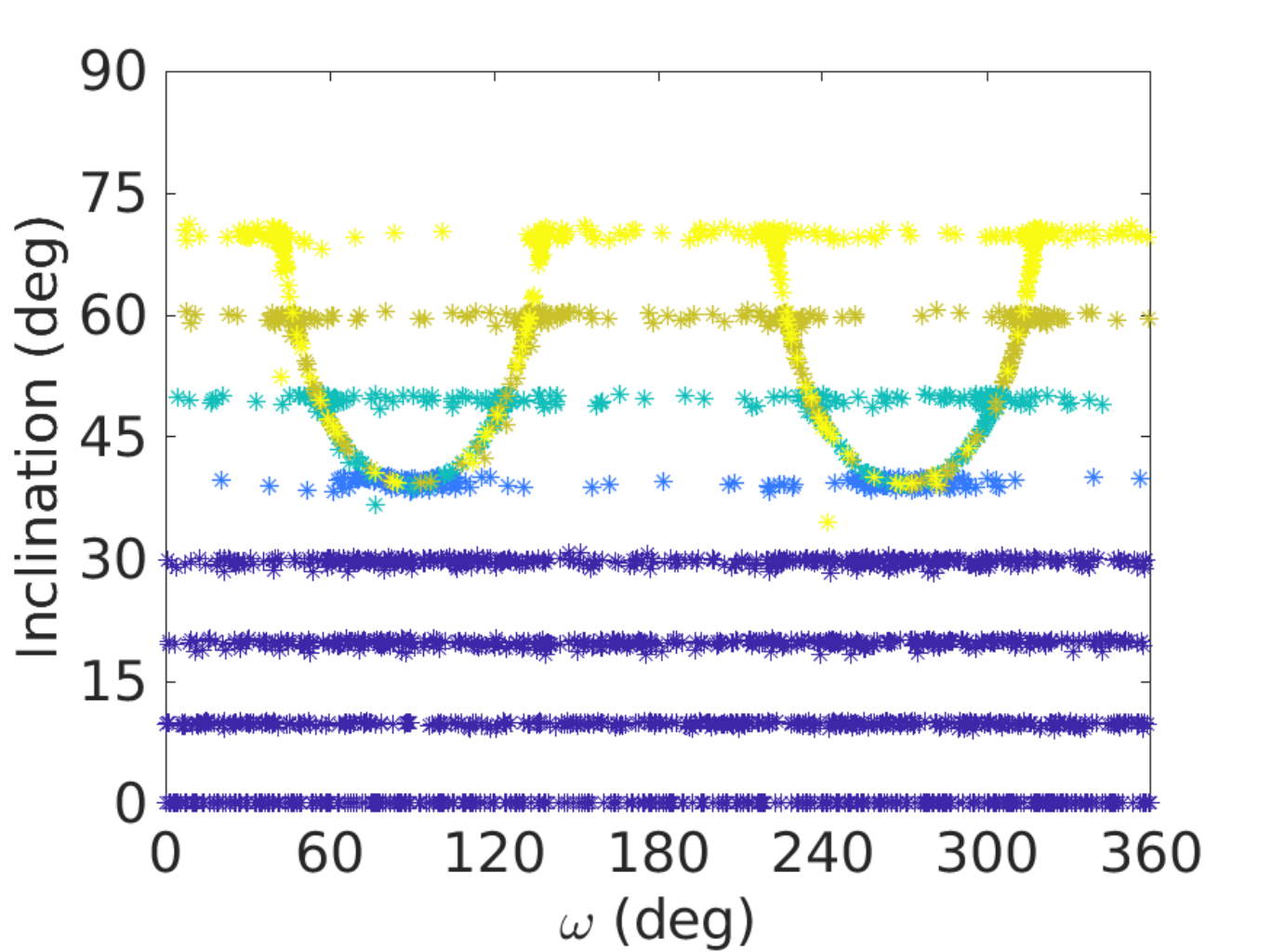}}}
        \subfloat{\resizebox{0.33\hsize}{!}{\includegraphics{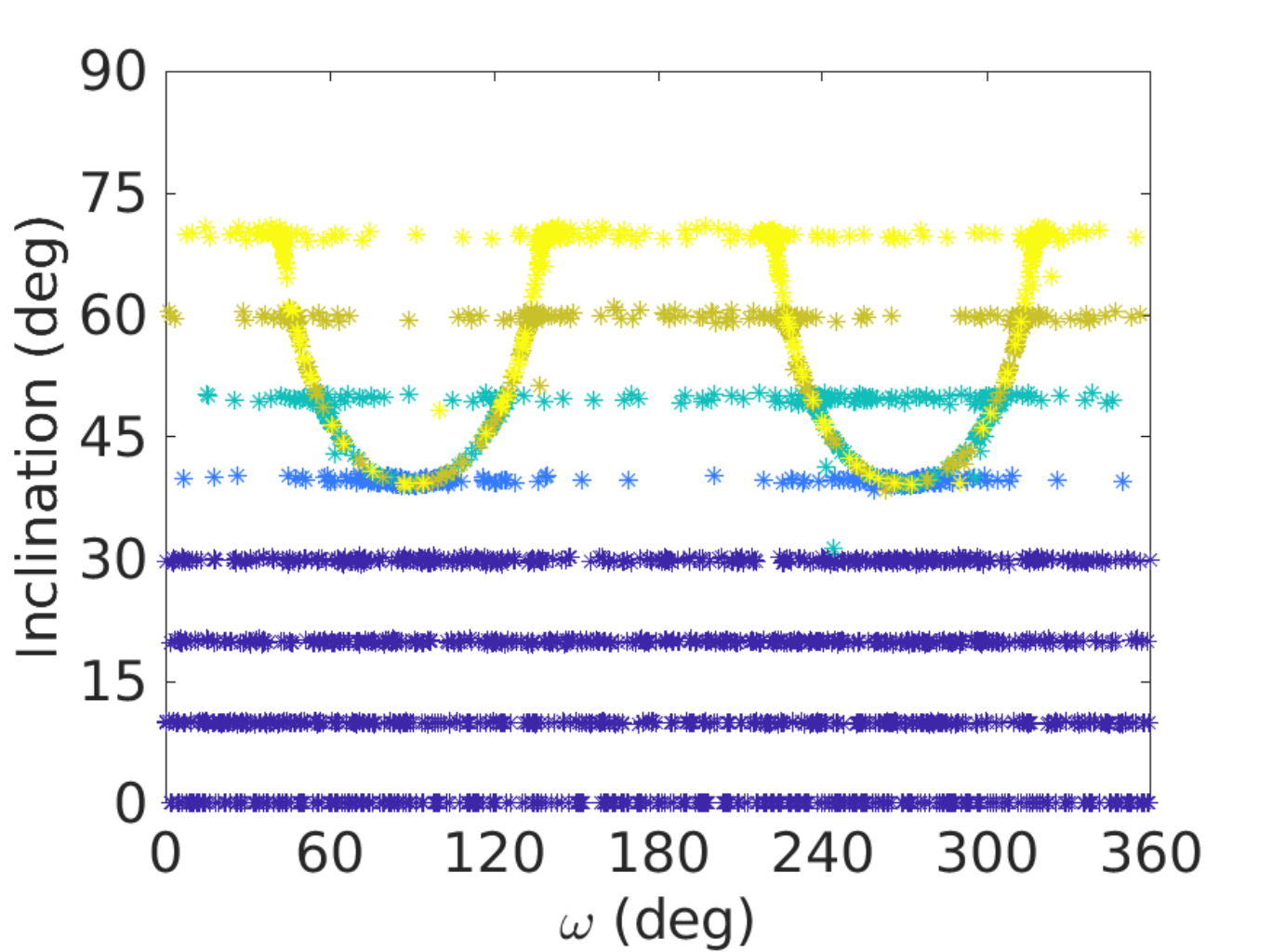}}}\\
        \subfloat{\resizebox{0.33\hsize}{!}{\includegraphics{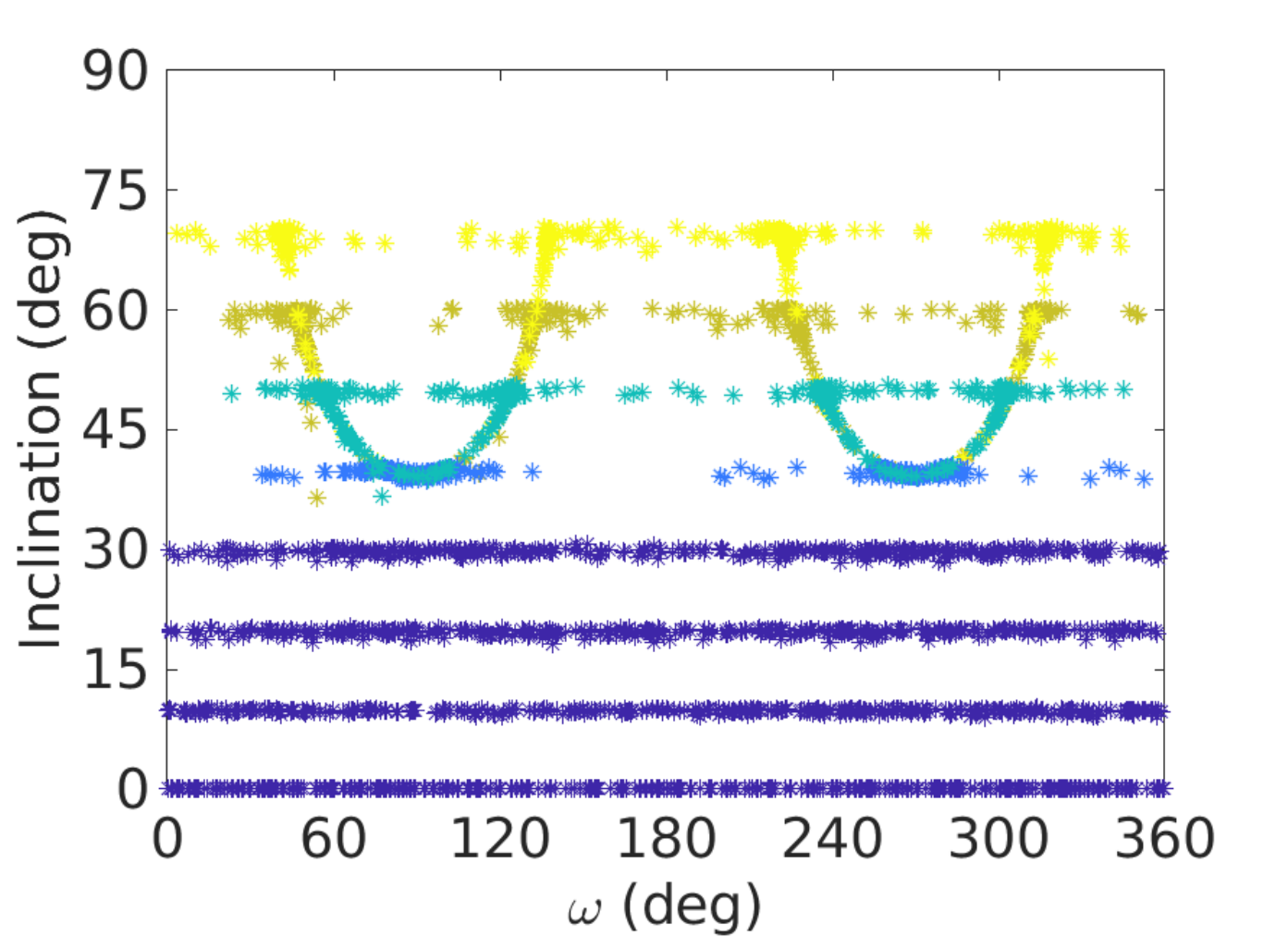}}}
        \subfloat{\resizebox{0.33\hsize}{!}{\includegraphics{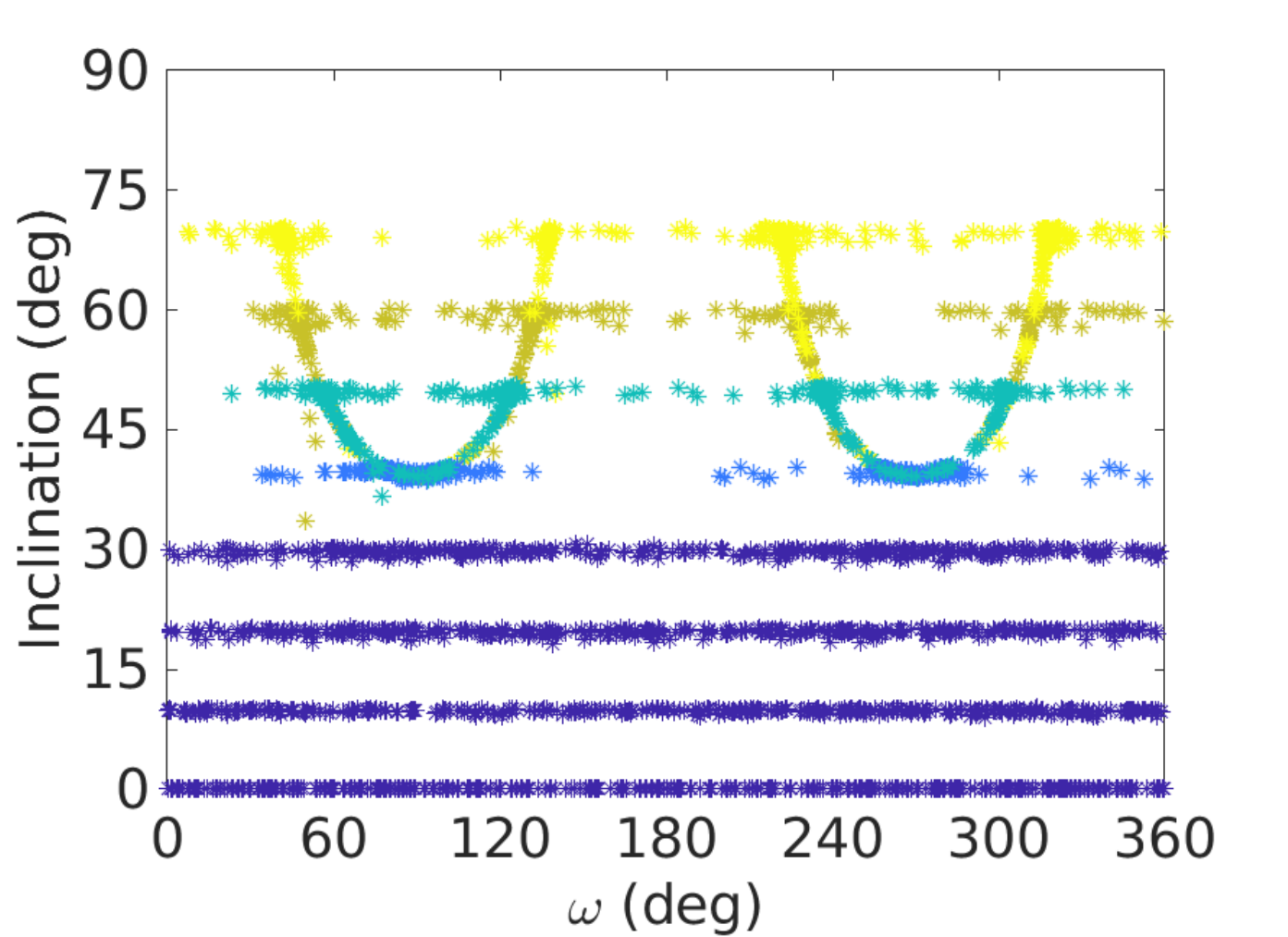}}}
        \subfloat{\resizebox{0.33\hsize}{!}{\includegraphics{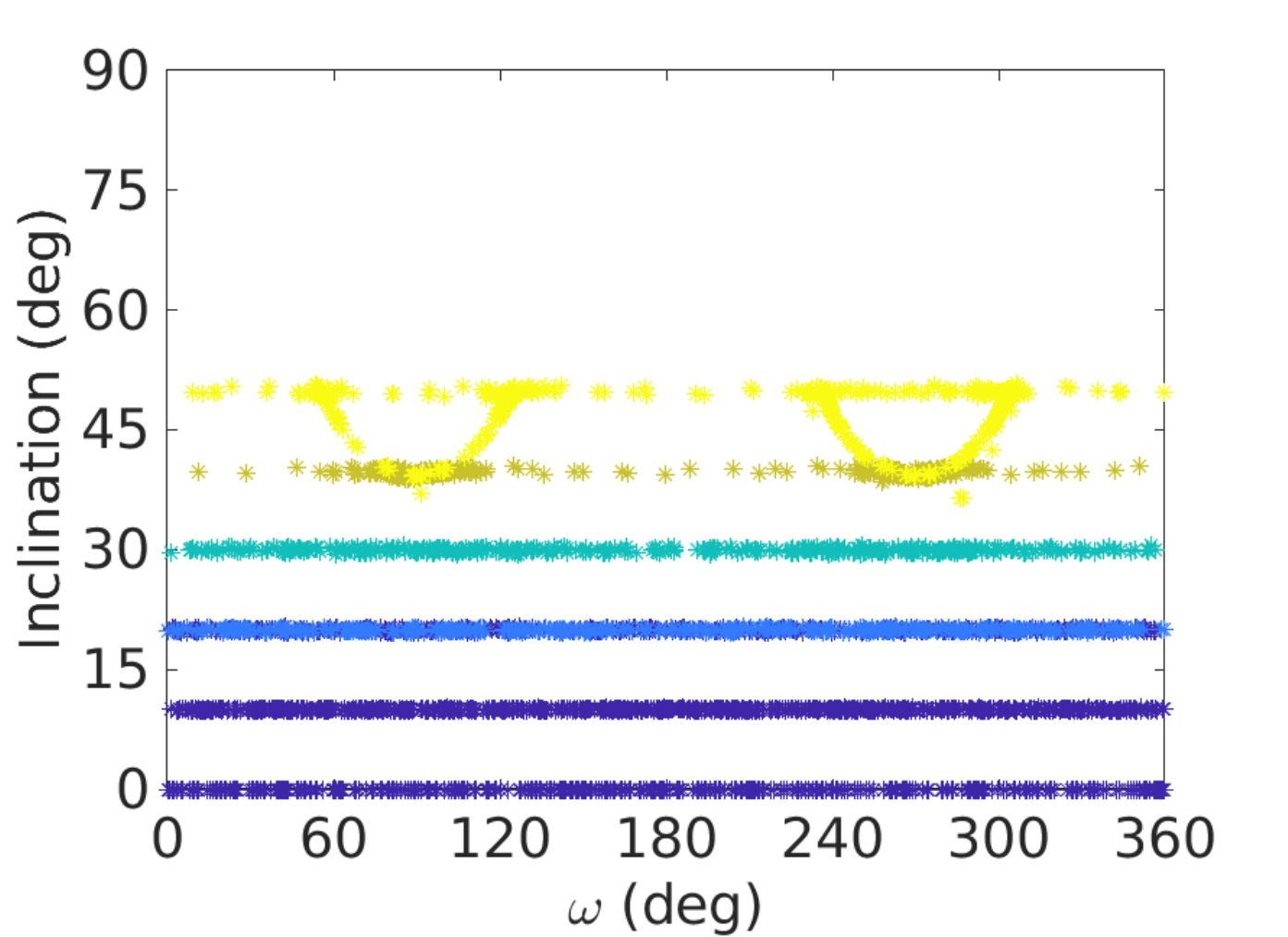}}}\\
        \center{\subfloat{\resizebox{0.6\hsize}{!}{\includegraphics{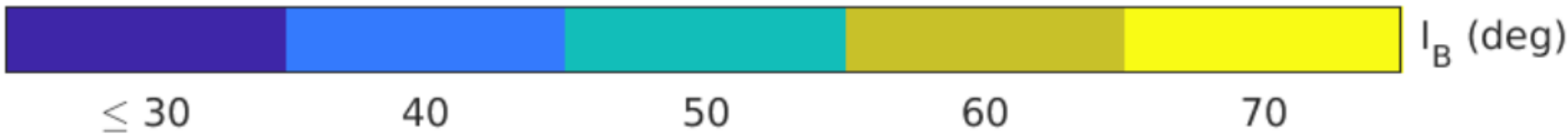}}}}
        \caption{Same as Fig.~\ref{i_w} for different migration and damping rates. Rates are shown unchanged in the left panels, Type II migration rate scaled by the factor $\cos({i_{pl}})$ indicated in the middle panels, and no eccentricity and inclination damping considered in the right panels. Two initial inclination values of the planet with respect to the disk plane are also shown ($i_{pl}$ is fixed to $0^\circ$ in the top panels and $20^\circ$ in the bottom panels).}
        \label{test_selfgravity}
\end{figure*}

Finally, we show that the trend observed in the right panel of Fig.~\ref{incl} can be explained by the conservation of the Kozai constant (green curves). Indeed, for a given value of the Kozai constant, the ranges of possible eccentricity and inclination values for the planet are fixed by the relation $h=\sqrt{1-e_{pl}^2}\cos i_{pl}$. Here the Kozai constant is evaluated at the dispersal of the disk and has been approximated by the quantity $\cos i_{B}$, since the disk tends to circularize the planetary orbit and to keep it in the disk through eccentricity and inclination damping.

\section{Discussion and conclusions}
\label{Discussion}

In this work, we studied the dynamical influence of a wide binary companion on the migration of a single giant planet in the protoplanetary disk. The simulations were made using the SyMBA symplectic n-body integrator modified for binary star systems and eccentricity and inclination damping formulae for the disk influence. By varying the eccentricity and inclination of the binary companion as well as the physical and orbital parameters of the planet, we carried out 3200 numerical simulations and carefully analyzed the orbital configuration of the planets after 100~Myr.
   
As a consequence of the Lidov-Kozai mechanism, the influence of the wide binary companion is strong for highly inclined companions even during the disk phase. We showed that the orbital parameters of the planets at the end of the simulations gather around two arc-shaped curves in the planes $(\omega_{pl}, e_{pl})$. We noticed that only about 36$\%$ of the planets with a highly inclined binary companion ($i_B \ge 40^\circ$) end up in a Lidov-Kozai resonant state (with the libration of the pericenter argument). Nevertheless, the nonresonant evolutions are strongly influenced by the Lidov-Kozai resonance and present high eccentricity and inclination variations associated with a circulation around the Lidov-Kozai islands. Using an analytical quadrupolar Hamiltonian approach, we realized phase portraits in which the different dynamical evolutions can easily be observed.

It is important to note that in this work we neglected the influence of the companion star on the disk, in particular, the induced nodal precession of the disk. To study the robustness of our results, and in particular the impact that nodal precession could have, we ran several tests by considering different migration rates and eccentricity and inclination damping rates as well as a location of the planet initially outside of the disk. The results of these tests are summarized in Fig.~\ref{test_selfgravity} which shows, for the different tests, the same analysis as in Fig.~\ref{i_w}, namely, the planetary inclination as a function of the argument of the pericenter at the end of the simulation, in which the color code indicates the initial inclination of the binary companion $i_B$. In the top panels, the planets are initially inside of the disk plane ($i_{pl}=0^\circ$), while the planets have an initial inclination of $i_{pl}=20^\circ$ with respect to the disk plane in the bottom panels. Three modifications for the planetary migration are considered. In the left panels, the migration rates and eccentricity and inclinations damping rates are unchanged. In the middle panels, the Type II migration rate is scaled by the factor $\cos({i_{pl}})$. In the right panels, no eccentricity and inclination damping is included. Thus, in total, six tests were realized, each one consisting in 3200 simulations, as previously detailed. As observed in Fig.~\ref{test_selfgravity}, the modifications performed have no significant impact on the results and the arc-shaped curves are present in each test. 

These extra simulations make us confident in the robustness of the results presented in this work. We obtained similar outcomes when changing the planetary inclination and, therefore, the mutual inclination between the planet and the disk. Therefore, we believe that the addition of the nodal precession of the disk caused by the binary star, inducing a change in the mutual inclination, does not significantly modify the long-term evolution of the planets.

For this study, we did not include additional effects due to the disk and the binary that are likely to have a significant impact on the evolution of the planet. Aside from the nodal precession of the disk caused by the binary star, we did not consider the aspidal and nodal precessions caused by the gravitational potential of the disk, which could dominate over the precession induced by the binary star (e.g., \cite{Zanazzi_2018b}). Also, the general relativity and the stellar oblateness could compete or suppress the Lidov-Kozai effect. The influence of these additional effects on our results is left for future work.

We believe that since it is currently estimated that about half of the Sun-like stars are part of multiple-star systems, the complete understanding of the influence of a wide binary companion on the planetary eccentricities and inclinations is very important to fully explain the observational parameter distributions of the exoplanets.

\begin{acknowledgements}
The authors would like to warmly thank J. Teyssandier, K. Tsiganis and A. Crida for useful discussions as well as the anonymous referee for his/her comments that improve the present document. The work of A. Roisin is supported by a F.R.S.-FNRS research fellowship. Computational resources have been provided by the Consortium des Equipements de Calcul Intensif, supported by the FNRS-FRFC, the Walloon Region, and the University of Namur (Conventions No. 2.5020.11, GEQ U.G006.15, 1610468 et RW/GEQ2016).
\end{acknowledgements}

\bibliographystyle{aa}
\bibliography{ArXiv}

\begin{thebibliography}{51}
\expandafter\ifx\csname natexlab\endcsname\relax\def\natexlab#1{#1}\fi

\bibitem[{{Artymowicz} \& {Lubow}(1994)}]{Artymowicz_1994}
{Artymowicz}, P. \& {Lubow}, S.~H. 1994, \apj, 421, 651

\bibitem[{Bataille {et~al.}(2018)Bataille, Libert, \& Correia}]{Bataille_2018}
Bataille, M., Libert, A.-S., \& Correia, A. C.~M. 2018, Monthly Notices of the
  Royal Astronomical Society, 479, 4749

\bibitem[{{Batygin} {et~al.}(2011){Batygin}, {Morbidelli}, \&
  {Tsiganis}}]{Batygin_2011}
{Batygin}, K., {Morbidelli}, A., \& {Tsiganis}, K. 2011, \aap, 533, A7

\bibitem[{{Bazs{\'o}} {et~al.}(2017){Bazs{\'o}}, {Pilat-Lohinger}, {Eggl},
  {Funk}, {Bancelin}, \& {Rau}}]{Bazso_2017}
{Bazs{\'o}}, {\'A}., {Pilat-Lohinger}, E., {Eggl}, S., {et~al.} 2017, \mnras,
  466, 1555

\bibitem[{{Bitsch} {et~al.}(2013){Bitsch}, {Crida}, {Libert}, \&
  {Lega}}]{Bitsch_2013}
{Bitsch}, B., {Crida}, A., {Libert}, A.-S., \& {Lega}, E. 2013, \aap, 555, A124

\bibitem[{{Bourbaki}(1972)}]{Bourbaki_1972}
{Bourbaki}, N. 1972, {El\'ements de Math\'ematiques: Groupes et Alg\`ebres de
  Lie, Hermann Ed. (Paris)}

\bibitem[{{Campbell} {et~al.}(1988){Campbell}, {Walker}, \&
  {Yang}}]{Campbell_1988}
{Campbell}, B., {Walker}, G.~A.~H., \& {Yang}, S. 1988, \apj, 331, 902

\bibitem[{Chambers {et~al.}(2002)Chambers, Quintana, Duncan, \&
  Lissauer}]{Chambers_2002}
Chambers, J.~E., Quintana, E.~V., Duncan, M.~J., \& Lissauer, J.~J. 2002, The
  Astronomical Journal, 123, 2884

\bibitem[{Crida \& Morbidelli(2007)}]{Crida_2007}
Crida, A. \& Morbidelli, A. 2007, Monthly Notices of the Royal Astronomical
  Society, 377, 1324

\bibitem[{{Duncan} {et~al.}(1998){Duncan}, {Levison}, \& {Lee}}]{Duncan_1998}
{Duncan}, M.~J., {Levison}, H.~F., \& {Lee}, M.~H. 1998, \aj, 116, 2067

\bibitem[{{Duquennoy} \& {Mayor}(1991)}]{Duquennoy_1991}
{Duquennoy}, A. \& {Mayor}, M. 1991, \aap, 248, 485

\bibitem[{Fabrycky \& Tremaine(2007)}]{Fabrycky_2007}
Fabrycky, D. \& Tremaine, S. 2007, The Astrophysical Journal, 669, 1298

\bibitem[{{Fu} {et~al.}(2015a){Fu}, {Lubow}, \& {Martin}}]{Fu_2015_II}
{Fu}, W., {Lubow}, S.~H., \& {Martin}, R.~G. 2015a, \apj, 807, 75

\bibitem[{{Goldreich} \& {Tremaine}(1979)}]{Goldreich_1979}
{Goldreich}, P. \& {Tremaine}, S. 1979, \apj, 233, 857

\bibitem[{{Haghighipour}(2010)}]{Haghighipour_2010}
{Haghighipour}, N., ed. 2010, Astrophysics and Space Science Library, Vol. 366,
  {Planets in Binary Star Systems}

\bibitem[{{Haghighipour} \& {Raymond}(2007)}]{Haghighipour_2007}
{Haghighipour}, N. \& {Raymond}, S.~N. 2007, \apj, 666, 436

\bibitem[{{Hatzes} {et~al.}(2003){Hatzes}, {Cochran}, {Endl}, {McArthur},
  {Paulson}, {Walker}, {Campbell}, \& {Yang}}]{Hatzes_2003}
{Hatzes}, A.~P., {Cochran}, W.~D., {Endl}, M., {et~al.} 2003, \apj, 599, 1383

\bibitem[{Holman \& Wiegert(1999)}]{Holman_1999}
Holman, M.~J. \& Wiegert, P.~A. 1999, The Astronomical Journal, 117, 621

\bibitem[{{Innanen} {et~al.}(1997){Innanen}, {Zheng}, {Mikkola}, \&
  {Valtonen}}]{Innanen_1997}
{Innanen}, K.~A., {Zheng}, J.~Q., {Mikkola}, S., \& {Valtonen}, M.~J. 1997,
  \aj, 113, 1915

\bibitem[{{Ivanov} {et~al.}(1999){Ivanov}, {Papaloizou}, \&
  {Polnarev}}]{Ivanov_1999}
{Ivanov}, P.~B., {Papaloizou}, J.~C.~B., \& {Polnarev}, A.~G. 1999, \mnras,
  307, 79

\bibitem[{{Kaib} {et~al.}(2013){Kaib}, {Raymond}, \& {Duncan}}]{Kaib_2013}
{Kaib}, N.~A., {Raymond}, S.~N., \& {Duncan}, M. 2013, \nat, 493, 381

\bibitem[{{Kley} \& {Haghighipour}(2014)}]{Kley_2014}
{Kley}, W. \& {Haghighipour}, N. 2014, \aap, 564, A72

\bibitem[{{Kozai}(1962)}]{Kozai_1962}
{Kozai}, Y. 1962, \aj, 67, 591

\bibitem[{{Kraus} \& {Ireland}(2012)}]{Kraus_2012}
{Kraus}, A.~L. \& {Ireland}, M.~J. 2012, \apj, 745, 5

\bibitem[{Laskar \& Robutel(2001)}]{Laskar_2001}
Laskar, J. \& Robutel, P. 2001, Celestial Mechanics and Dynamical Astronomy,
  80, 39

\bibitem[{Lee \& Peale(2002)}]{Lee_2002}
Lee, M.~H. \& Peale, S.~J. 2002, The Astrophysical Journal, 567, 596

\bibitem[{{Libert} \& {Tsiganis}(2009)}]{Libert_2009}
{Libert}, A.-S. \& {Tsiganis}, K. 2009, \mnras, 400, 1373

\bibitem[{{Lidov}(1962)}]{Lidov_1962}
{Lidov}, M.~L. 1962, \planss, 9, 719

\bibitem[{{Lissauer}(1993)}]{Lissauer_1993}
{Lissauer}, J.~J. 1993, \araa, 31, 129

\bibitem[{{Lubow} \& {Martin}(2016)}]{Lubow_2016}
{Lubow}, S.~H. \& {Martin}, R.~G. 2016, \apj, 817, 30

\bibitem[{{Martin} {et~al.}(2016){Martin}, {Lubow}, {Nixon}, \&
  {Armitage}}]{Martin_2016}
{Martin}, R.~G., {Lubow}, S.~H., {Nixon}, C., \& {Armitage}, P.~J. 2016,
  \mnras, 458, 4345

\bibitem[{{Martin} {et~al.}(2014){Martin}, {Nixon}, {Lubow}, {Armitage},
  {Price}, {Do{\u g}an}, \& {King}}]{Martin_2014}
{Martin}, R.~G., {Nixon}, C., {Lubow}, S.~H., {et~al.} 2014, \apjl, 792, L33

\bibitem[{{Marzari} {et~al.}(2005){Marzari}, {Weidenschilling}, {Barbieri}, \&
  {Granata}}]{Marzari_2005}
{Marzari}, F., {Weidenschilling}, S.~J., {Barbieri}, M., \& {Granata}, V. 2005,
  \apj, 618, 502

\bibitem[{{Matsumoto} {et~al.}(2012){Matsumoto}, {Nagasawa}, \&
  {Ida}}]{Matsumoto_2012}
{Matsumoto}, Y., {Nagasawa}, M., \& {Ida}, S. 2012, \icarus, 221, 624

\bibitem[{Nelson {et~al.}(2000)Nelson, Papaloizou, Masset, \&
  Kley}]{Nelson_2000}
Nelson, R.~P., Papaloizou, J. C.~B., Masset, F., \& Kley, W. 2000, Monthly
  Notices of the Royal Astronomical Society, 318, 18

\bibitem[{Papaloizou \& Larwood(2000)}]{Papaloizou_2000}
Papaloizou, J. C.~B. \& Larwood, J.~D. 2000, Monthly Notices of the Royal
  Astronomical Society, 315, 823

\bibitem[{{Picogna} \& {Marzari}(2015)}]{Picogna_2015}
{Picogna}, G. \& {Marzari}, F. 2015, \aap, 583, A133

\bibitem[{{Queloz} {et~al.}(2000){Queloz}, {Mayor}, {Weber}, {Bl{\'e}cha},
  {Burnet}, {Confino}, {Naef}, {Pepe}, {Santos}, \& {Udry}}]{Queloz_2010}
{Queloz}, D., {Mayor}, M., {Weber}, L., {et~al.} 2000, \aap, 354, 99

\bibitem[{Quintana {et~al.}(2007)Quintana, Adams, Lissauer, \&
  Chambers}]{Quintana_2007}
Quintana, E.~V., Adams, F.~C., Lissauer, J.~J., \& Chambers, J.~E. 2007, The
  Astrophysical Journal, 660, 807

\bibitem[{{Rabl} \& {Dvorak}(1988)}]{Dvorak_1988}
{Rabl}, G. \& {Dvorak}, R. 1988, \aap, 191, 385

\bibitem[{{Rafikov} \& {Silsbee}(2015)}]{Rafikov_2015}
{Rafikov}, R.~R. \& {Silsbee}, K. 2015, \apj, 798, 69

\bibitem[{{Raghavan} {et~al.}(2010){Raghavan}, {McAlister}, {Henry}, {Latham},
  {Marcy}, {Mason}, {Gies}, {White}, \& {ten Brummelaar}}]{Raghavan_2010}
{Raghavan}, D., {McAlister}, H.~A., {Henry}, T.~J., {et~al.} 2010, The
  Astrophysical Journal Supplement, 190, 1

\bibitem[{Savonije {et~al.}(1994)Savonije, Papaloizou, \& Lin}]{Savonije_1994}
Savonije, G.~J., Papaloizou, J. C.~B., \& Lin, D. N.~C. 1994, Monthly Notices
  of the Royal Astronomical Society, 268, 13

\bibitem[{{Schwarz} {et~al.}(2016){Schwarz}, {Funk}, {Zechner}, \&
  {Bazs{\'o}}}]{Schwarz_2016}
{Schwarz}, R., {Funk}, B., {Zechner}, R., \& {Bazs{\'o}}, {\'A}. 2016, \mnras,
  460, 3598

\bibitem[{{Shakura} \& {Sunyaev}(1973)}]{Shakura_1973}
{Shakura}, N.~I. \& {Sunyaev}, R.~A. 1973, \aap, 24, 337

\bibitem[{{Sotiriadis} {et~al.}(2017){Sotiriadis}, {Libert}, {Bitsch}, \&
  {Crida}}]{Sotiriadis_2017}
{Sotiriadis}, S., {Libert}, A.-S., {Bitsch}, B., \& {Crida}, A. 2017, \aap,
  598, A70

\bibitem[{{Thebault} \& {Haghighipour}(2015)}]{Thebault_2016}
{Thebault}, P. \& {Haghighipour}, N. 2015, {Planet Formation in Binaries},
  309--340

\bibitem[{{Th{\'e}bault} {et~al.}(2006){Th{\'e}bault}, {Marzari}, \&
  {Scholl}}]{Thebault_2006}
{Th{\'e}bault}, P., {Marzari}, F., \& {Scholl}, H. 2006, \icarus, 183, 193

\bibitem[{{Xiang-Gruess} \& {Papaloizou}(2014)}]{Xiang_2014}
{Xiang-Gruess}, M. \& {Papaloizou}, J.~C.~B. 2014, \mnras, 440, 1179

\bibitem[{{Zanazzi} \& {Lai}(2017)}]{Zanazzi_2017a}
{Zanazzi}, J.~J. \& {Lai}, D. 2017, \mnras, 467, 1957

\bibitem[{{Zanazzi} \& {Lai}(2018)}]{Zanazzi_2018b}
{Zanazzi}, J.~J. \& {Lai}, D. 2018, \mnras, 478, 835

\end{thebibliography}

\end{document}